\documentclass[lettersize,journal]{IEEEtran}

\usepackage{cite}
\usepackage[table]{xcolor}
\usepackage{amsmath,amssymb,amsfonts}
\usepackage{algorithm}
\usepackage{algorithmic}
\usepackage{graphicx}
\usepackage{textcomp}
\usepackage{subcaption}
\usepackage{xcolor}
\usepackage{tikz}
\usepackage{makecell}
\usepackage{multirow}

\renewcommand{\thetable}{\arabic{table}} 

\begin{document}

\title{Trustless Privacy-Preserving Data Aggregation on Ethereum with Hypercube Network Topology}

\author{\IEEEauthorblockN{Goshgar Ismayilov, Can Ozturan} \\
\IEEEauthorblockA{\textit{Computer Engineering Department} \\
\textit{Bogazici University}\\
Istanbul, Turkey}}

\maketitle

\begin{abstract}
The privacy-preserving data aggregation is a critical problem for many applications where multiple parties need to collaborate with each other privately to arrive at certain results. Blockchain, as a database shared across the network, provides an underlying platform on which such aggregations can be carried out with a decentralized manner. Therefore, in this paper, we have proposed a scalable privacy-preserving data aggregation protocol for summation on the Ethereum blockchain by integrating several cryptographic primitives including commitment scheme, asymmetric encryption and zero-knowledge proof along with the hypercube network topology. The protocol consists of four stages as \textit{contract deployment}, \textit{user registration}, \textit{private submission} and \textit{proof verification}. The analysis of the protocol is made with respect to two main perspectives as security and scalability including computational, communicational and storage overheads. In the paper, the zero-knowledge proof, smart contract and web user interface models for the protocol are provided. We have performed an experimental study in order to identify the required gas costs per individual and per system. The general formulation is provided to characterize the changes in gas costs for the increasing number of users. The zero-knowledge proof generation and verification times are also measured.
\end{abstract}


\section{Introduction}

\IEEEPARstart{B}{lockchain} is fundamentally a decentralized ledger technology that has gained significant attention in recent years due to its high potential to revolutionize a wide range of industries. For the first time, it was proposed in 2008 in a white paper with the title of "Bitcoin: A Peer-to-Peer Electronic Cash System" in order to introduce a linked data structure that allows for the creation of a secure, transparent, and immutable records of transactions  \cite{nakamoto2008}. This distributed architecture of blockchain makes it resistant to external tampering and fraud, which leads to a viable and attractive solution for various real-world applications including financial transactions \cite{schar2021}, supply chain management \cite{vara2018}, energy management \cite{ali2020}, e-voting \cite{yavuz2018} and bartering \cite{ozturan2020}. Despite all these revolutionary achievements, there exist numerous issues still needed to be addressed including privacy-preserving protocols while maintaining the transparency and security of the blockchain at the same time.

Zero-knowledge proof is among the commonly used tools in cryptography to address such privacy concerns. In a standard zero-knowledge proof scheme, a prover aims to convince a verifier about the correctness of a statement without disclosing the statement itself. Following its first proposal in the literature \cite{goldwasser1989}, several number-theoretic and graph-theoretic approaches have been developed to build zero-knowledge proofs. The discrete logarithm, integer factorization and quadratic residuocity problems are the number-theoretic approaches \cite{cramer1994, feige1988, guillou1988, fiat1986} while the graph-3 colorability and graph (non-)isomorphism problems are the graph-theoretic approaches \cite{bellare1990, goldreich1991, mohr2007} on which zero-knowledge proof scheme can be built. However, after the advancements of the blockchain technologies, more complex protocols have been proposed including zk-SNARKs (Zero-Knowledge Succinct Non-Interactive Argument of Knowledge) \cite{sasson2014}, zk-STARKs (Zero-Knowledge Scalable Transparent Argument of Knowledge) \cite{sasson2018}, Bulletproofs \cite{bunz2019} and Plonk \cite{gabizon2019}. In the recent years, these protocols have been applied to numerous problems in the blockchain literature successfully \cite{gabay2020, li2020, jeong2021, li2021}. In our work, we have integrated zk-SNARKs through the Zokrates toolbox \cite{eberhardt2018} to generate and verify zero-knowledge proofs for the correctness of the aggregations.

In the realm of blockchain technologies, privacy-preserving data aggregation is a critical issue to be addressed. It simply refers to the procedures of collecting and combining data from multiple blockchain parties while still maintaining their privacy. In the literature, there are several approaches to provide such procedures including secure multi-party computation, differential privacy and homomorphic encryption \cite{lindell2005, peng2019, singh2021}. There are also many applications fields of privacy-preserving data aggregation protocols in blockchain including healthcare to share data about the patients while still maintaining their privacy \cite{almalki2021}, finance to combine data of investors to identify potential risks and frauds without violating their rights \cite{fan2020}, marketing to understand general consumer behaviour and preferences \cite{mao2018}, e-voting to count the votes to determine the winner \cite{mols2020} and governmental activities to improve public services \cite{drogkaris2015}. This work proposes a general-purpose privacy-preserving data aggregation protocol for summation based on the decentralized nature of blockchain and the public verifiability of zero-knowledge proof.

The motivation of our work is to enable crowds to create large-scale groups in order to perform the data aggregation privately, collectively, without any trusted party and scalable in terms of the number of parties. To the best of our knowledge, our paper is among the pioneering works in the literature to support these significant properties during the data aggregation on blockchain. Along with the summation operation, the protocol is economical and scalable to be extended into different mathematical operations in the future including the multiplication, the average and the vector-based calculations. It has also potential in terms of the number of real-world applications including (i) e-voting to count the number of votes per candidate and (ii) privacy-preserving bartering to exchange a set of tokens in return for another set of tokens where data aggregation is necessary to determine whether the total set of tokens supplied to the system is higher the total set of tokens demanded from the system \cite{ozturan2020}. 

Our work presents a novel technique to the literature by integrating the hypercube network topology into the verifiable zero-knowledge proof protocol in the scope of blockchain. With the help of the hypercube networks, different parties are enabled to be properly coordinated for aggregation with a decentralized manner. There exist several challenges to be overcome during the design and the development of the resulting protocol. These challenges include (i) \textit{trustless} as the elimination of any trusted third party at any stage of the aggregation process, (ii) \textit{coordination} between the multiple parties at each dimension of the hypercube networks and (iii) \textit{privacy} as the verification of the secret value aggregations while still not revealing them. The approaches to overcome these challenges will be discussed in the next sections throughout the work.

The main contributions of this paper can be outlined as follows:
\begin{itemize}
\item A private data aggregation protocol for summation has been proposed on Ethereum using cryptographic primitives (including the commitment, public-key cryptography and zero-knowledge proof schemes) and hypercube network topology. For the protocol, the zero-knowledge proof, the smart contract and the web user interface models are designed.

\item The proposed protocol has been analyzed and compared with respect to two different perspectives including security and scalability with computation, communication, storage overheads. 

\item An extensive experimental study has been carried out where the Ethereum gas cost and the zero-knowledge proof generation times have been considered as our performance measures. The problem requirements satisfied through the proposed protocol have been also identified. Furthermore, the gas cost has been formulated in order to characterize the changes from the individual user and whole protocol perspectives for a high number of users.

\end{itemize}

The remainder of the paper is organized as follows. Section 2 reviews the related work while Section 3 gives preliminary information for cryptographic primitives and hypercube network topology. Section 4 proposes the private data aggregation protocol and their details. Section 5 analyzes the protocol from security and scalability perspectives. Section 6 presents the experimental study and discusses the results with Ethereum gas costs and proof generation times. Finally, Section 7 concludes the paper.

\section{Related Work}

In the literature, privacy-preserving data aggregation protocols have been extensively studied  \cite{damgard2012, damgard2013, baldimtsi2015, bonawitz2017, balle2020, ranbaduge2020}. Especially, with the advancements of the blockchain technologies in the recent years, the protocols initially developed for traditional systems as well as the novel protocols have been successfully integrated to the smart contract-based applications \cite{wang2021, mouris2021, mols2020, singh2021, fan2020}. In the following work \cite{wang2021}, a traditional protocol for privacy-preserving data aggregation, SPDZ \cite{damgard2012}, has been utilized to provide energy storage sharing in blockchain where the parties can compute total energy demand without revealing their own private energy demands. The work consists of four stages as (i) \textit{initialization} to choose system parameters, (ii) \textit{pre-operation scheduling} to compute total demands via SPDZ, (iii) \textit{cost-sharing operation} to split the cost with respect to the individual demands and (iv) \textit{post-operation} to sign individual energy storage service receipts. In SPDZ, a secret value is basically divided into multiple values and they are distributed to different parties. The construction of the same value requires the collaboration of these parties. Along with SPDZ, the work combines non-interactive zero-knowledge summation proof, cryptographic commitment and public-key cryptography in order to eliminate trusted setup. However, the main limitation of the work is that the computational and the communicational costs linearly increase (i.e. in non-scalable fashion) with an increasing number of parties where gas costs are exceeded for higher than 25 parties. The growth rate should be smaller in order to have a scalable system, which is done in our work.

Another work \cite{mouris2021} has proposed a general-purpose privacy-preserving data aggregation protocol on the Ethereum blockchain for different statistical operations including summation, mean and histograms. The work has three different actors as (i) participants with private values, (ii) curator who is responsible for collecting and computing the encrypted summation through homomorphic commitments and (iii) analyst who encrypts the final commitment and verifies the correctness of the operations performed. The protocol in the work has four stages as (i) \textit{key-generation} where the analyst generates a homomorphic keys to distribute the public key with the participants, (ii) \textit{encrypt-prove-commit} where each participant encrypts the private value with that public key and commits the commitment by generating a zero-knowledge proof, (iii) \textit{aggregation} where the curator verifies the proofs and aggregates the commitments to obtain the final commitment and (iv) the analyst receives the final commitment and decrypts it through the private key to learn the result of the operation. The work combines homomorphic encryption to encrypt secret values, zero-knowledge proof for the correctness of operations and special heuristic to convert interactive zero-knowledge proof to non-interactive zero-knowledge proof. However, the work suffers from two crucial issues where the first issue is that it is largely dependent upon trusted third party, curator. The second issue is that it has not been passed through a peer review process.

E-voting can be also discussed from the privacy-preserving data aggregation perspective where the individual votes must be tallied to determine the winner of the election. In the following work \cite{mols2020}, an e-voting application has been proposed for the Ethereum blockchain based on three collaboratively running smart contracts. The first contract is \textit{registration authority} which is responsible for registering or unregistering voters allowed to participate in the elections. The second contract is \textit{election factory} where a new election instance can be created by specifying the start and the finish time. Finally, the third contract is \textit{election} where there are three stages as \textit{before-election} to add candidates, \textit{during-election} to submit encrypted votes to the system and \textit{after-election} to count and sum the votes. The work combines homomorphic commitment and zero-knowledge proof along with a front-end application for usability. However, the implementation of zero-knowledge proof has not been completed due to technical limitations.

The work \cite{liu2023} focuses on the privacy-preserving data aggregation and model training problem for federated learning of internet-of-things (IoT) devices.  It proposes a hierarchical approach where private data are progressively aggregated from the local nodes to the cluster models, to the edge models and finally to the global models. The work also presents an aggregation node selection protocol (i.e. \textit{DDPG (Deep Deterministic Policy Gradient)}) in order to select the most optimal set of nodes for the aggregation. However, the work suffers from the existence of the central authorities (i.e. leaders) during the data aggregation and the result verification stages. Meanwhile, it has a lack of experiments for blockchain-based performance in terms of gas cost. When taking all these works into consideration, it is said that our protocol enables large-scale and general-purpose data aggregation for summation  privately, collectively, distributedly and without any central authority on the blockchain environment.  

\section{Preliminary}

In the work, the public information (i.e. secret values of the users) is made private through the replacement of that information with its corresponding commitment. In order to be able to verify the correctness of the commitments and their pairwise aggregations, the zero-knowledge proof scheme is used. At each stage of the hypercube network topology, the paired users are confidentially communicated by using the public-key cryptography technique. Hence, we will review the commitment scheme, the public-key cryptography, the zero-knowledge proof scheme and finally the hypercube network topology without loss of generality in this section. The commonly used notations are also given in Table 1.

\begin{table}[htbp]
\caption{Notations}
\footnotesize
\begin{center}
\begin{tabular}{|l|l|}
\hline
\textbf{Notation} & \textbf{Description} \\
\hline
$Comm$ & Commit function  \\
\hline
$Vfy$ & Verification function  \\
\hline
$Enc$ & Encryption function  \\
\hline
$Dec$ & Decryption function  \\
\hline
$ZkpGen$ & Proof generation function  \\
\hline
$ZkpVfy$ & Proof verification function  \\
\hline
$KeyGen$ & Key-pair generation function  \\
\hline
$Prng$ & Pseudo-random number generation function \\
\hline
$\pi$ & Proof  \\
\hline
$r$ & Salt value  \\
\hline
$N$ & Number of users in a group  \\
\hline
$b'$ & Bit representation \\
\hline
$h'$ & Hex representation  \\
\hline
$\oplus$ & xor function \\
\hline
\end{tabular}
\end{center}
\footnotesize
\end{table}

\subsection{Commitment Scheme}

A commitment scheme is a one-way cryptographic hash function which maps a value to another value (i.e. commitment) \cite{menezes2018}. This operation fundamentally allows a party to commit a secret value without revealing it to the other parties. A commitment scheme has three main phases as \textit{setup}, \textit{commit} and \textit{verify}. In the first phase, the parties agree upon the same system parameters. In the second phase, a party computes the commitment from a secret value $x$ and a salt value to mask $r$ as $c = Comm(x, r)$. In the third phase, other parties can verify the correctness of the commitment after the secret value is revealed, $b = Vfy(c, x, r)$ where $b$ is Boolean.

A commitment scheme must satisfy two properties in the most simple form as \textit{hiding} (i.e. \textit{preimage resistance}) and \textit{binding} (i.e. \textit{2nd-preimage resistance}) \cite{menezes2018}. The binding property enforces that there must be no two values whose commitments are the same so that a party cannot change the pre-committed secret value later, $ Pr[x_1 \neq x_2 \mid  Comm(x_1, r_1) = Comm(x_2, r_2)] < \epsilon $. On the other hand, the hiding property requires that the commitment itself must not reveal any information about the secret value to the other parties, $  \mid Pr[c_1 = Comm(x_1, r_1) \mid c_1 ] -  Pr[c_2 = Comm(x_2, r_2) \mid c_2 ]  \mid < \epsilon  $. In general, the commitment schemes are considered as suitable cryptographic primitives to construct zero-knowledge proofs. For the proposed protocol of this paper, the SHA-256 commitment scheme \cite{sha256} from the family of \textit{S}ecure \textit{H}ash \textit{A}lgorithms is selected to be used due to its built-in support of Zokrates cryptographic toolbox \cite{eberhardt2018}. 

\subsection{Public-Key Cryptography}

A public-key cryptographic (i.e. asymmetric cryptography) scheme is a trap-door one-way cryptographic function which is used for both data encryption and digital signature \cite{menezes2018}. It solves the key distribution and management issue of symmetric cryptographic schemes by introducing a pair of distinct keys as public key and private key for each party, $(pk, sk)$. While a public key $pk$ can be distributed among the other parties, a private key $sk$ must not be shared. On the other hand, unlike in a commitment scheme, it is easy to compute in one direction but difficult to compute in another direction without any trap-door information in a public-key encryption scheme.

A public-key cryptographic scheme for data encryption has three different phases as \textit{key generation}, \textit{encryption} and \textit{decryption} \cite{menezes2018}. In the first phase, each party generates a pair of public and private keys. In the second phase, each party encrypts a secret message through the public key $pk$ of the recipient, $ E = Enc(m, pk) $. Finally, in the third phase, each party can decrypt the secret message through its own private key $ sk $, $ m = D = Dec(E, sk)$. In this paper, the ECIES (\textit{E}lliptic \textit{C}urve \textit{I}ntegrated \textit{E}ncryption \textit{S}cheme) \cite{ecies} is selected for data encryption, which combines the elliptic curve cryptography with asymmetric encryption cryptography. It is used to construct direct private channels between the parties of the proposed protocol on the public Ethereum blockchain where the public keys are stored in the smart contract.

\subsection{Zero-knowledge Proof Scheme}

Zero-knowledge proof scheme was first proposed by Goldwasser, Micali and Rackoff in 1985 \cite{goldwasser1989} where a prover party aims to convince another verifier party about the knowledge of a secret value without revealing the secret value itself. The proof generation and proof verification are two main responsibilities of the prover and the verifier, respectively. Given a language $L$ on which proof scheme is built and two parties as prover $P$ and verifier $V$, it is considered to be a zero-knowledge proof in case the following three properties are satisfied:

\begin{itemize}
\item \textit{Completeness}: A prover can convince a honest verifier about the knowledge of secret with very high probability if the claim is correct, $ x \in L \to Pr[(P, V)(x) \neq 1] < \epsilon$.

\item \textit{Soundness}: A cheating prover can convince a honest verifier about the knowledge of secret with very negligible probability if the claim is not correct, $ x \notin L \to Pr[(P, V)(x) = 1] < \epsilon$.

\item \textit{Zero-Knowledge Proof}: A verifier can learn no additional information other than the correctness of the claim after the interaction with a prover. A scheme is considered to be zero-knowledge if there exists a simulator algorithm for every verifier $ V^* $ where any information learned after the interaction with a prover can also be learned through that simulator, $ (P, V^*)(x) \approx SM^*(x) $.

\end{itemize}

In this paper, the non-interactive zero-knowledge of commitment scheme is used where a prover party convinces other verifier parties about the knowledge of a secret value through the correctness of the commitment \cite{blum1991}. There exists three phases of this scheme as \textit{setup}, \textit{proof generation} and \textit{proof verification}. In the first phase, the prover commits a secret value and stores the commitment in the smart contract, $c = Comm(x, r)$. In the second phase, the prover generates the proof using its own secret value and the commitment, $\pi = ZkpGen(x, r, c)$. In the third and the final phase, the verifier can check the correctness of the proof without any interaction with the prover, $ b = ZkpVfy(\pi, c) $, where $b$ is a Boolean value.

Zokrates is a privacy-preserving cryptographic toolbox in order to implement the verifiable zero-knowledge proofs on Ethereum \cite{eberhardt2018}. Specifically, it supports a family of zero-knowledge proofs, zk-SNARKs (Zero-Knowledge Succinct Non-Interactive Arguments of Knowledge) \cite{sasson2014}, which is \textit{succinct} in terms of proof size; \textit{non-interactive} with no requirement of prover-verifier interaction; and also \textit{zero-knowledge} by fulfilling the third property of zero-knowledge proofs. There exist two main advantages of Zokrates \cite{eberhardt2018}. Firstly, it allows to generate proofs off-chain and later verify them on-chain in order to prevent the high gas cost of Ethereum. Secondly, it provides a platform where high-level proofs can be easily implemented without any consideration of complex cryptographic and mathematical operations. Zokrates is used in this paper to generate zero-knowledge proofs in the browser and to verify these proofs in the smart contracts.

\begin{figure*}[htbp]
\begin{subfigure}{0.30\textwidth}
  \centering
  \hspace*{0cm}
  \includegraphics[width=1\linewidth]{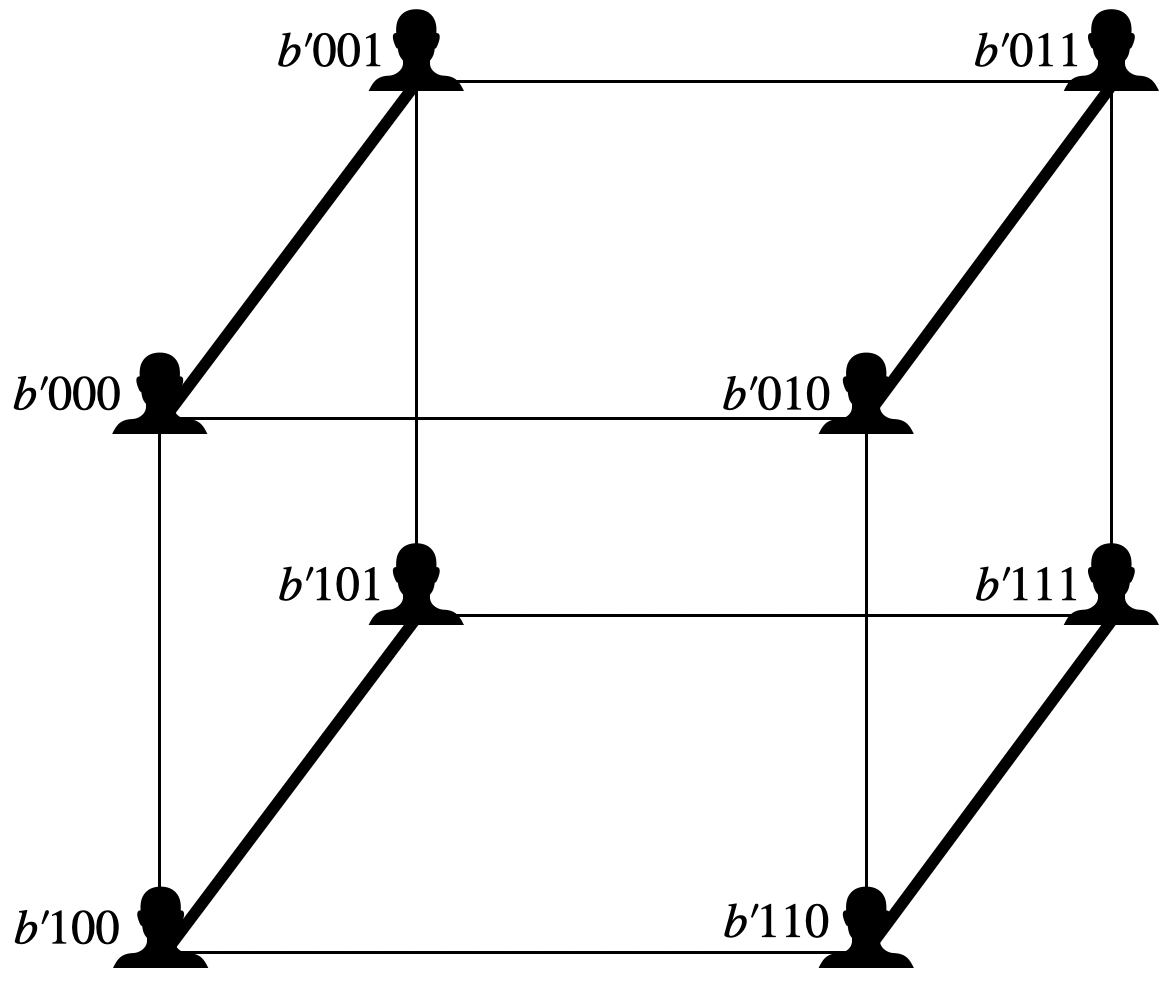}
  \caption{1st Communication Dimension}
\end{subfigure}
\begin{subfigure}{0.30\textwidth}
  \centering
  \hspace*{0cm}
  \includegraphics[width=1\linewidth]{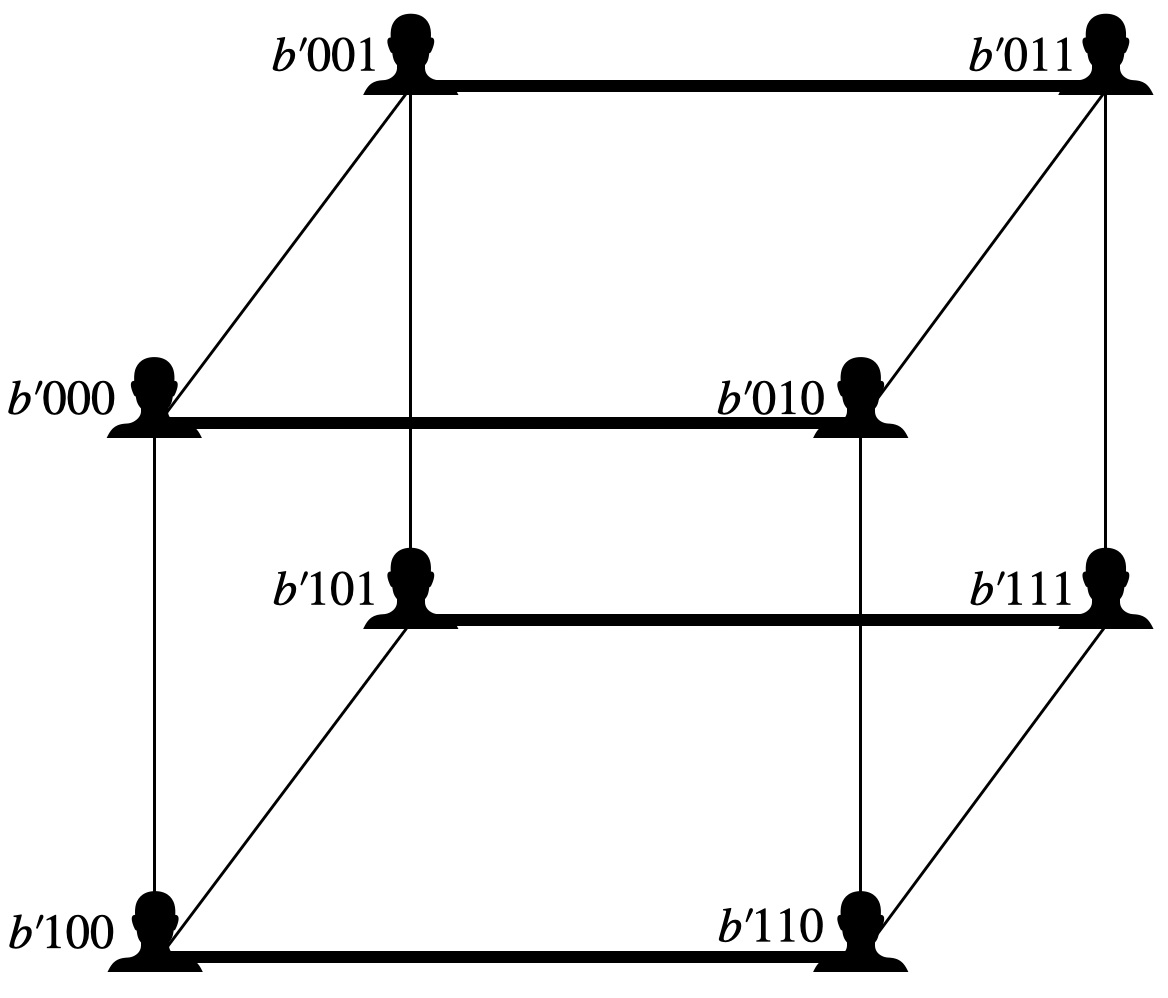}
  \caption{2nd Communication Dimension}
\end{subfigure}
\begin{subfigure}{0.30\textwidth}
  \centering
  \hspace*{0cm}
  \includegraphics[width=1\linewidth]{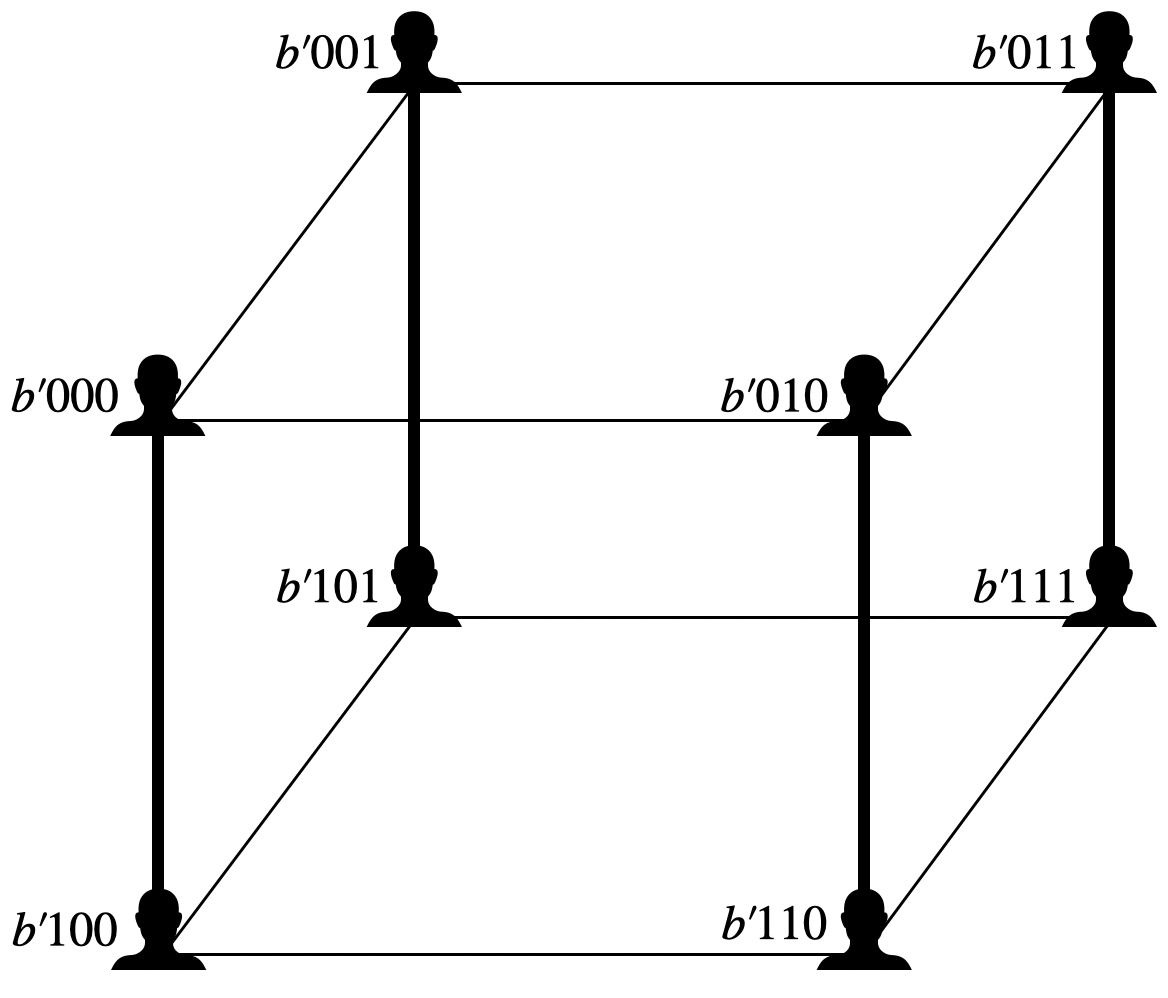}
  \caption{3rd Communication Dimension}
\end{subfigure}
\caption{Hypercube Diagrams for 1st, 2nd and 3rd Communication Flows, respectively}
\end{figure*}

\subsection{Hypercube Topology}

A network topology, in the simplest of terms, refers to the arrangement of the nodes and their respective connections in a communication network. The hypercube network topology is among these topologies, which can be defined as a closed, convex and multi-dimensional structure to connect the nodes through bi-directional links to perform specific sort of inter-communication \cite{leighton2014}. Each topology has its own unique trade-offs and has potential to affect the performance of the protocols running over that topology to a certain extent. One of the main advantages of the hypercube topology is its high degree of communication scalability for a growing number of nodes. For instance, only $log(N)$ parallel communications are required for a hypercube network for $N$ number of nodes \cite{leighton2014}.

Given a graph network $G=(V, E)$, $V$ is the set of vertices referring to different parties in the blockchain while $E$ is the set of edges referring to the direct communication links between these parties. Each node $u$ is represented by a distinct bit string of length $log(|V|)$ as $u = u_{log(|V|)-1}\ldots  u_j\ldots u_0$ where $u_j \in \{0, 1\}$. A graph network is called a hypercube network if and only if the vertices whose bit strings differ in exactly one position are connected to each other, $\mid u - u'  \mid = 1 $ where $\mid . \mid $ is the Hamming distance operator. In this paper, the hypercube structure is used as a logical layout in order to provide an underlying communication network to perform secure multi-party computation over the existing blockchain. For multiple parties, three resulting successive communication paths are shown in Fig. 1.

\section{Private Data Aggregation Protocol with Hypercube Topology}

In this section, we will define the privacy-preserving data aggregation problem by extracting its requirements. For the given problem, a cryptographic protocol is proposed including the general architecture, the zero-knowledge proof model, the smart contract model and the web user interface model. Especially, the user interface is the proof-of-concept implementation to show the correctness and the reproducibility of our entire system.

\subsection{Problem Definition and Requirements}

The privacy-preserving data aggregation for summation is a problem where a group of honest-but-curious parties are involved through a smart contract in Ethereum blockchain in order to perform summation over their secret values without disclosing them. There exist two main actors in the threat model as (i) the honest-but-curious parties which have consents for the data aggregation and the proof generation about the correctness of the off-chain calculations; and (ii) a smart contract deployed in a blockchain for the flow orchestration and management in order to arrive at the final result of the computation. The main assumptions of the threat model can be listed as follows: 

\begin{itemize}
\item The parties must follow the protocol until the end since the final aggregation will not be completed in the absence of a certain party rejecting to be involved with.

\item The parties must be honest-but-curious where they are also interested in learning as much information as possible about private data of the other parties.

\item The parties can collude with themselves except two parties communicating with a certain party in the first stages of two hypercube networks.

\item Blockchain as an underlying infrastructure must continue providing secure and immutable database functionality.

\end{itemize}

\begin{figure*}[htbp]
\centerline{\includegraphics[width=0.7\textwidth]{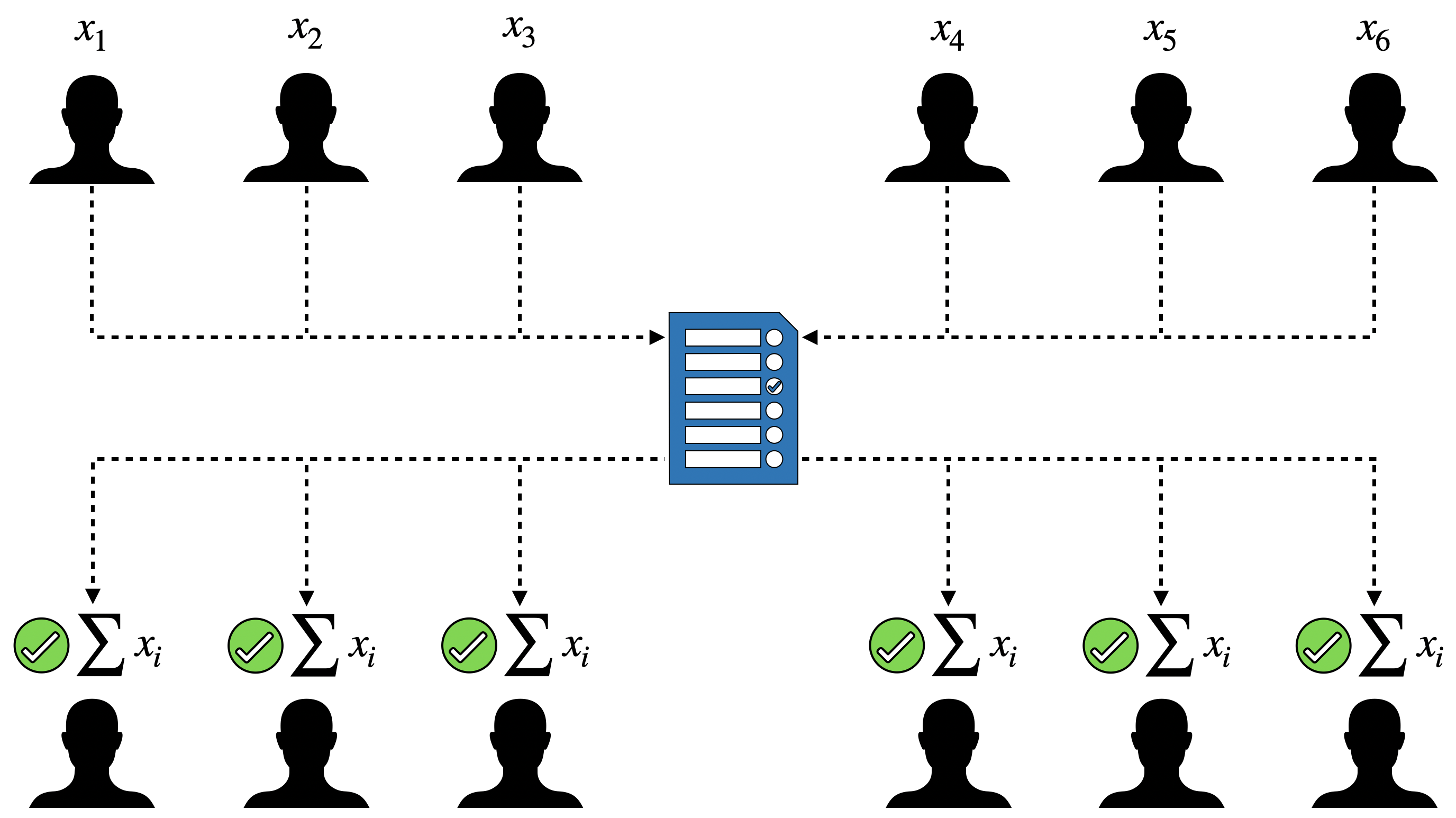}}
\caption{Verifiable Secure Multi-Party Computation over Smart Contract for Privacy-Preserving Data Aggregation for Summation}
\end{figure*}

Suppose there are $N$ parties where each party has a secret value $x_i$ that should not be shared to any other party in the system. The main goal of the privacy-preserving data aggregation problem for summation is to apply a summation function over a set of secret values ($x_0, x_1, ..., x_i, ..., x_{N-1}$). An illustrative figure for the privacy-preserving data aggregation problem is given in Fig. 2. The following problem requirements and design objectives are expected to be satisfied:

\begin{itemize}
\item \textit{Privacy}: any information about the secret values of the parties must not be revealed to other parties during the execution of the protocol.

\item \textit{Confidentiality}: any sensitive information must be protected from the unauthorized parties in direct communication channels between parties.

\item \textit{Trustless}: no party must rely on any trusted third party or any central organization on data aggregation.

\item \textit{Public Verifiability}: the correctness of any off-chain operation must be efficiently verifiable on-chain.

\item \textit{Authentication}: no other party must join a group of parties who perform privacy-preserving data aggregation if not authenticated earlier.

\item \textit{Non-Interactivity}: the protocol must allow the parties to be able to generate zero-knowledge proofs on their own without any interaction with the other parties.

\item \textit{Scalability}: the protocol must support a growing number of parties.

\item \textit{Correctness}: the output after the execution of the protocol must be correct with respect to the initial secret values and the function applied.

\item \textit{Usability}: the parties must interact with a user-friendly interface without having to know any complex cryptographic operations carried out in the protocol.
\end{itemize}

\subsection{Overall Architecture}

The proposed protocol has the following stages as (i) \textit{contract deployment} where a contract deployer deploys an instance of smart contract to blockchain to start an attempt for privacy-preserving summation for a group of parties, (ii) \textit{user registration} where each party registers to that group via address of the smart contract between the pre-specified start and finish times, (iii) \textit{private summation submission} where each party is paired with another peer party in each stage of the hypercube network to submit a private summation and finally (iv) \textit{proof verification} where each party must generate a zero-knowledge proof in order to verify the correctness of the private summation.

Once a contract deployer deploys a smart contract instance by specifying the registration start time and the registration time limit in terms of block time, the contract deployer should share the resulting contract creation address on blockchain to a group of parties in the \textit{contract deployment} stage. The protocol runs two parallel hypercube configurations as \textit{Configuration-A} and \textit{Configuration-B} where each party masks its own secret value with a random value. Each party must generate two commitments along with a key pair in the \textit{user registration} stage:

\begin{align}
& c^A = Comm(x + e, r_{x + e}) \\
& c^B = Comm(e, r_{e}) \\
& pk, sk = KeyGen(k) 
\end{align}

\noindent
where $c^A$ and $c^B$ are the resulting commitments of the commitment function $Comm$, based on the secret value $x$, the random value $e$ and their salting values $r_{x + e}$ and $r_{e}$ for \textit{Configuration-A} and \textit{Configuration-B}, respectively while $pk$ and $sk$ are the public and secret keys of the key generation function $KeyGen$ based on security parameter $k$. In the \textit{user registration} stage, the commitment and the public key will be submitted to the smart contract.

Although the distributed parties are physically connected via the underlying network protocol, they are logically connected to each other with respect to the hypercube network topology from the viewpoint of our protocol. The number of parties for privacy-preserving summation is required to compute the number of communications each party has during the execution of the protocol where it logarithmically increases with increasing number of parties. For every stage of the hypercube networks, the peer parties for a party is determined in the smart contract, based on the principle of $\mid u - u'  \mid = 1 $, as follows: 

\begin{align}
& u \oplus b'2^t = u^{{peer}^A} \\
& u \oplus b'2^{(log(N) - t - 1)} = u^{{peer}^B}
\end{align}

\noindent
where $u$ and $u^{peer}$ are the peer parties at a certain stage $t \in [0, log(N)-1]$. Note that the pairs of $u$ as $u^{{peer}^A}$ and $u^{{peer}^B}$ are different parties at the initial hypercube communication stage. While selecting pairs in these equations, \textit{Configuration-A} proceeds through the hypercube dimensions as $0, 1, 2, ..., log(N)-1$ while \textit{Configuration-B} uses the reverse dimensions from $log(N)-1$ to $0$. As an alternative approach, it is also possible for \textit{Configuration-B} to use the shifted dimensions $1, 2, ..., log(N)-1, 0$ as $u \oplus b'2^{(t + 1) \; mod \; log(N)} = u^{{peer}^B}$. This guarantees the pairs to be different at the initial communication stage as well. In the \textit{private summation submission} stage, the public keys of the peer parties are fetched to encrypt cumulative summation:

\begin{align}
& E^A = Enc([x^A_{sum} = x^A_{sum} + x^{{peer}^A}_{sum}], r^A, pk^{{peer}^A}) \\
& E^B = Enc([x^B_{sum} = x^B_{sum} + x^{{peer}^B}_{sum}], r^B, pk^{{peer}^B}) 
\end{align}

\noindent
where $E^A$ and $E^B$ are the resulting encryptions of the encryption function $Enc$ based on the cumulative summations of the party $x^A_{sum}$ and $x^B_{sum}$, the cumulative summations of the pairs of that party $x^{{peer}^A}_{sum}$ and $x^{{peer}^B}_{sum}$, the salting values $r^A$ and $r^B$ and finally the public keys of the pairs $pk^{{peer}^A}$ and $pk^{{peer}^A}$ for \textit{Configuration-A} and \textit{Configuration-B}, respectively. An illustration for two hypercube networks of the second protocol is given in Fig. 3. 

In the \textit{proof verification} stage, two encryptions are fetched from the smart contract for decryption as follows: 

\begin{align}
& [x^{{peer}^A}_{sum}, r^{{peer}^A}] = D^A = Dec([E^A], sk) \\
& [x^{{peer}^B}_{sum}, r^{{peer}^B}] = D^B = Dec([E^B], sk) 
\end{align}

\noindent
where $D^A$ and $D^B$ are the resulting decryptions of the decryption function $Dec$ based on the secret key of the party $sk$ for \textit{Configuration-A} and \textit{Configuration-B}, respectively. Each party perform cumulative summation operations separately for Configuration A ($x + e + x^{{peer}^A}_{sum}$) and Configuration B ($e + x^{{peer}^B}$). To prove the correctness of these two off-chain summation operations, two zero-knowledge proofs must be generated as follows: 

\begin{align}
\pi^A = ZkpGen([& x^A_{sum} = x^A_{sum} + x^{{peer}^A}_{sum}], \nonumber \\
& [r^A, r^{{peer}^A}, r^{{new}^A}], \nonumber \\
& [c^A, c^{{peer}^A}, c^{{new}^A}]) \\
\pi^B = ZkpGen([& x^B_{sum} = x^B_{sum} + x^{{peer}^B}_{sum}], \nonumber \\
& [r^B, r^{{peer}^B}, r^{{new}^B}], \nonumber \\
& [c^B, c^{{peer}^B}, c^{{new}^B}])
\end{align}

\noindent
where $\pi^A$ and $\pi^B$ are the resulting zero-knowledge proofs of the proof generation function $ZkpGen$ based on the current salting values of the party and its pairs as $r^A$, $r^B$, $r^{{peer}^A}$ and $r^{{peer}^B}$; and the new salting values of the party after summation $r^{{new}^A}$ and $r^{{new}^B}$; the current commitments of the party and its pairs as $c^A$, $c^B$, $c^{{peer}^A}$ and $c^{{peer}^B}$; and the new commitments of the party $c^{{new}^A}$ and $c^{{new}^B}$ for \textit{Configuration-A} and \textit{Configuration-B}, respectively. Note in these equations that from $x^A_{sum}$, $x^{{peer}^A}_{sum}$, $x^B_{sum}$ and $x^{{peer}^B}_{sum}$ at the previous hypercube communication stage, a new $x^A_{sum}$ and $x^B_{sum}$ are computed at the current communication stage, respectively. The equality sign here is used as assignment operation.

The resulting proofs are later submitted to the smart contract for verification: 

\begin{align}
[b^A , b^B] = ZkpVfy([& \pi^A, \pi^B, \nonumber \\
& c^A, c^{{peer}^A}, c^{{new}^A}, \nonumber \\
& c^B, c^{{peer}^B}, c^{{new}^B}])
\end{align}

\noindent
where $b^A$ and $b^B$ are the resulting Boolean values as true or false of the proof verification function $ZkpVfy$ based on the proofs $\pi^A$ and $\pi^B$ for \textit{Configuration-A} and \textit{Configuration-B}, respectively. If the proofs are correctly verified, the new commitments for the cumulative summation values are stored in the smart contract. The \textit{private summation submission} and \textit{the proof verification} are repetitively performed with respect to the number of dimensions in both hypercube networks. At the end, each party obtains the final result which is $\sum (x) = \sum (x + e) - \sum (e)$. The general flow of the second protocol is compactly presented in Fig. 4. An example calculations with two hypercubes for the second protocol is provided in Fig. 5 as well.

\begin{figure*}[htbp]
\begin{subfigure}{0.54\textwidth}
  \centering
  \includegraphics[width=1\linewidth]{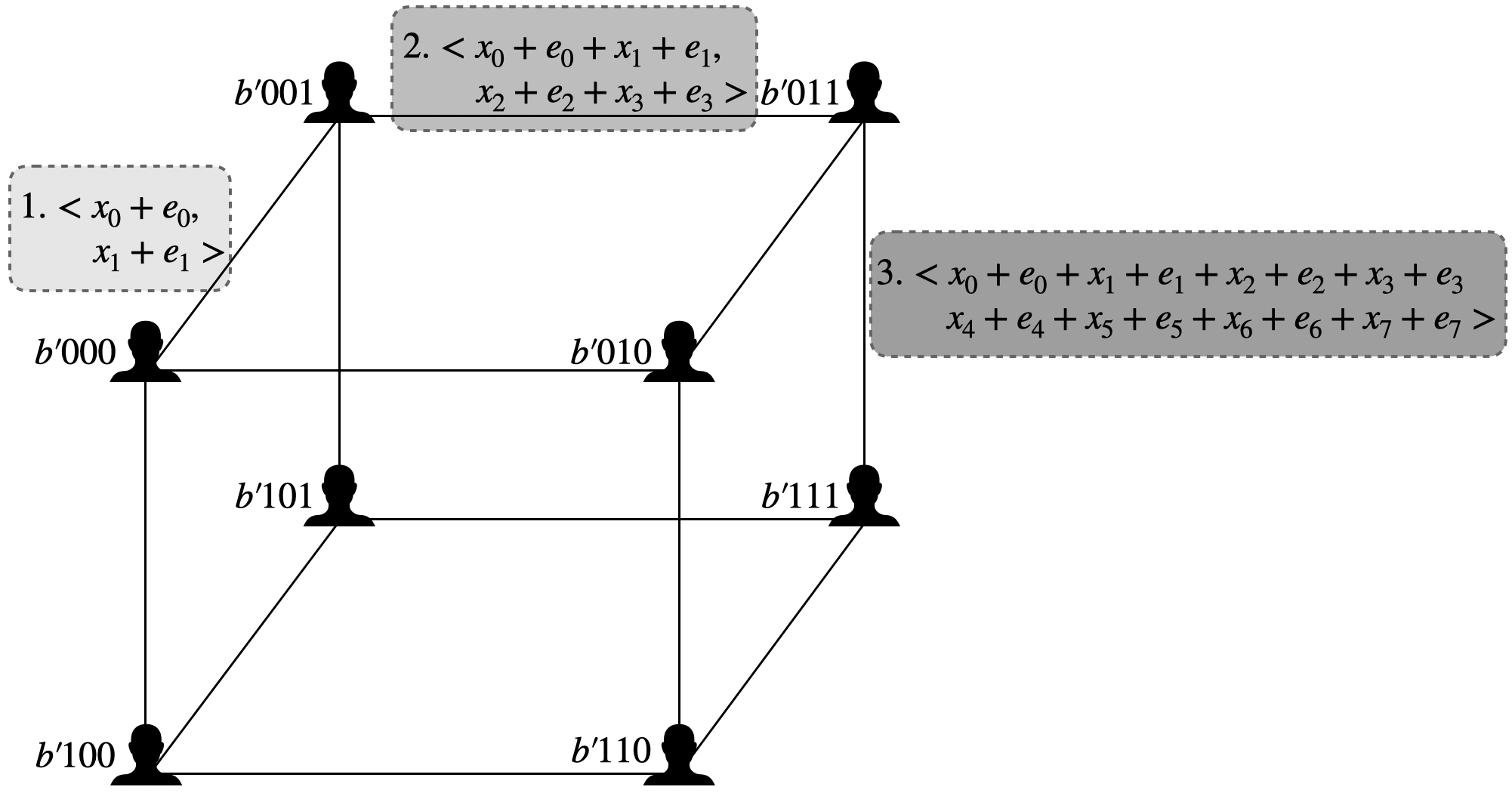}
  \caption{Hypercube-A Configuration}
\end{subfigure}
\begin{subfigure}{0.43\textwidth}
  \centering
  \includegraphics[width=1\linewidth]{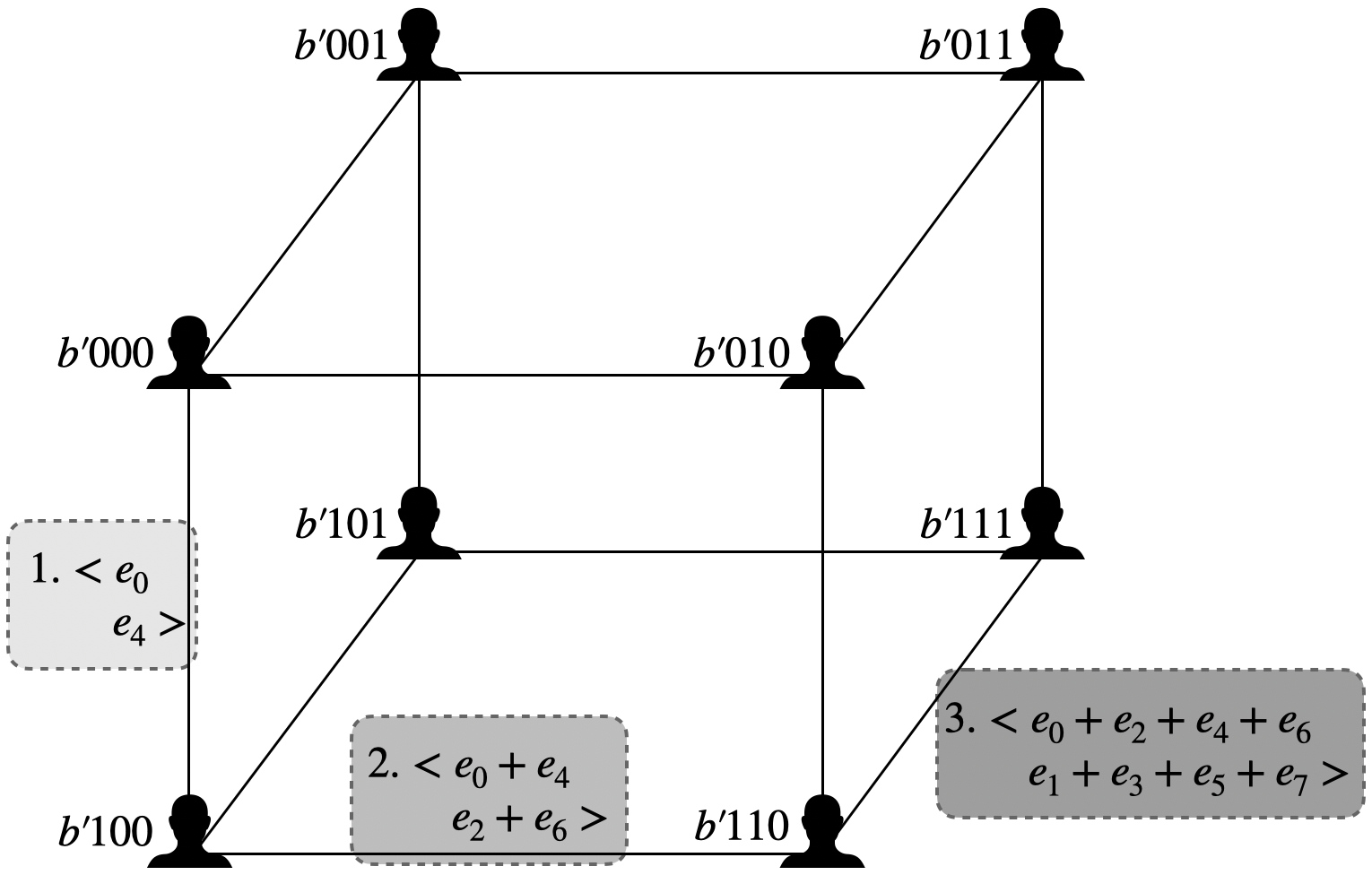}
  \caption{Hypercube-B Configuration}
\end{subfigure}
\caption{Hypercube Diagrams for Protocol 2 with 8 Parties in 3 Communication Flows}
\end{figure*}

\begin{figure*}[htbp]
\centerline{\includegraphics[width=1.0\textwidth]{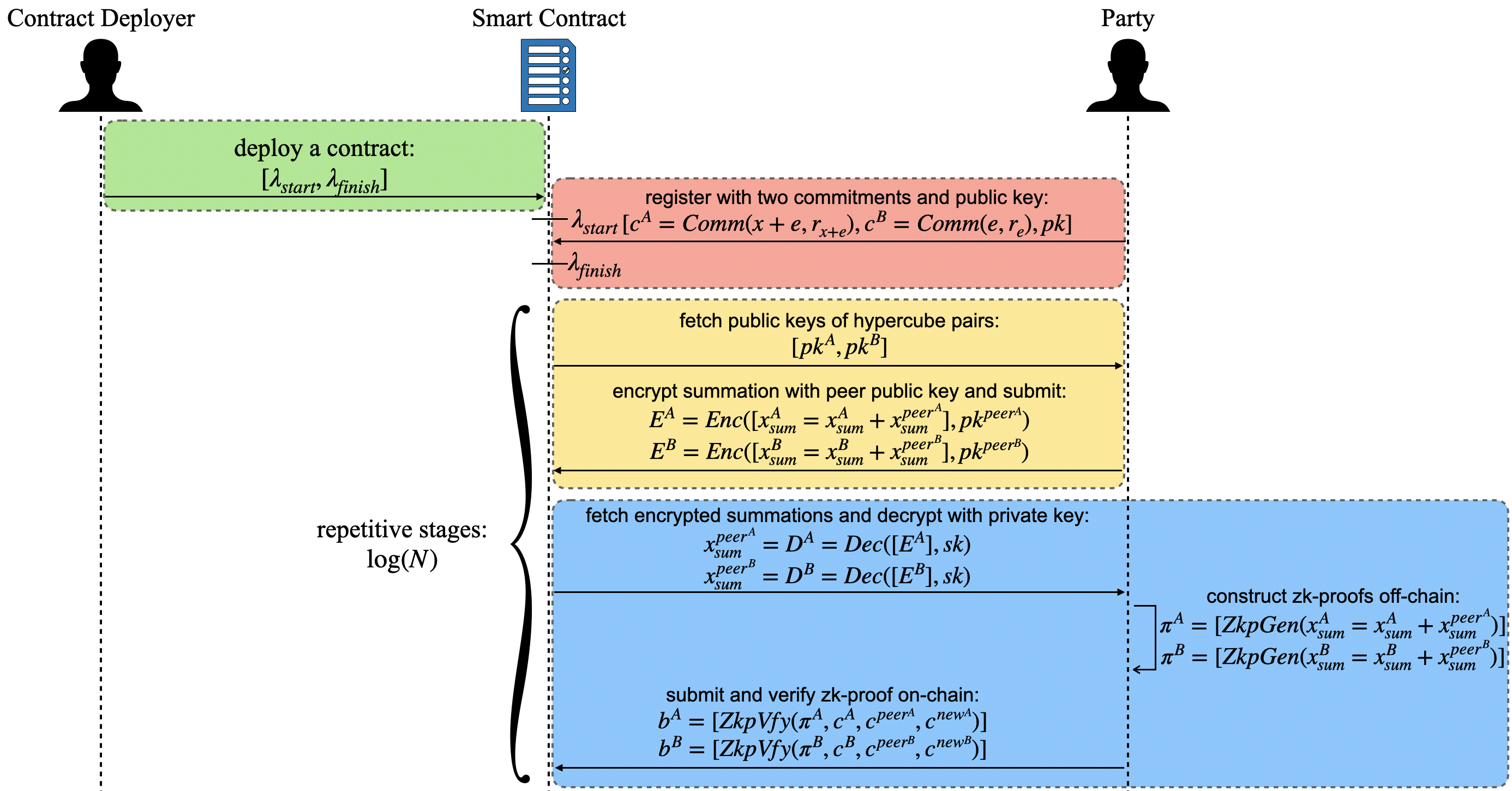}}
\caption{Sequence Diagram for Protocol 2 with Hypercube-A and -B Configurations: \textit{Contract Deployment}, \textit{User Registration}, \textit{Private Summation Submission} and \textit{Proof Verification} Stages}
\end{figure*}

\begin{figure*}[htbp]
\centerline{\includegraphics[width=1.0\textwidth]{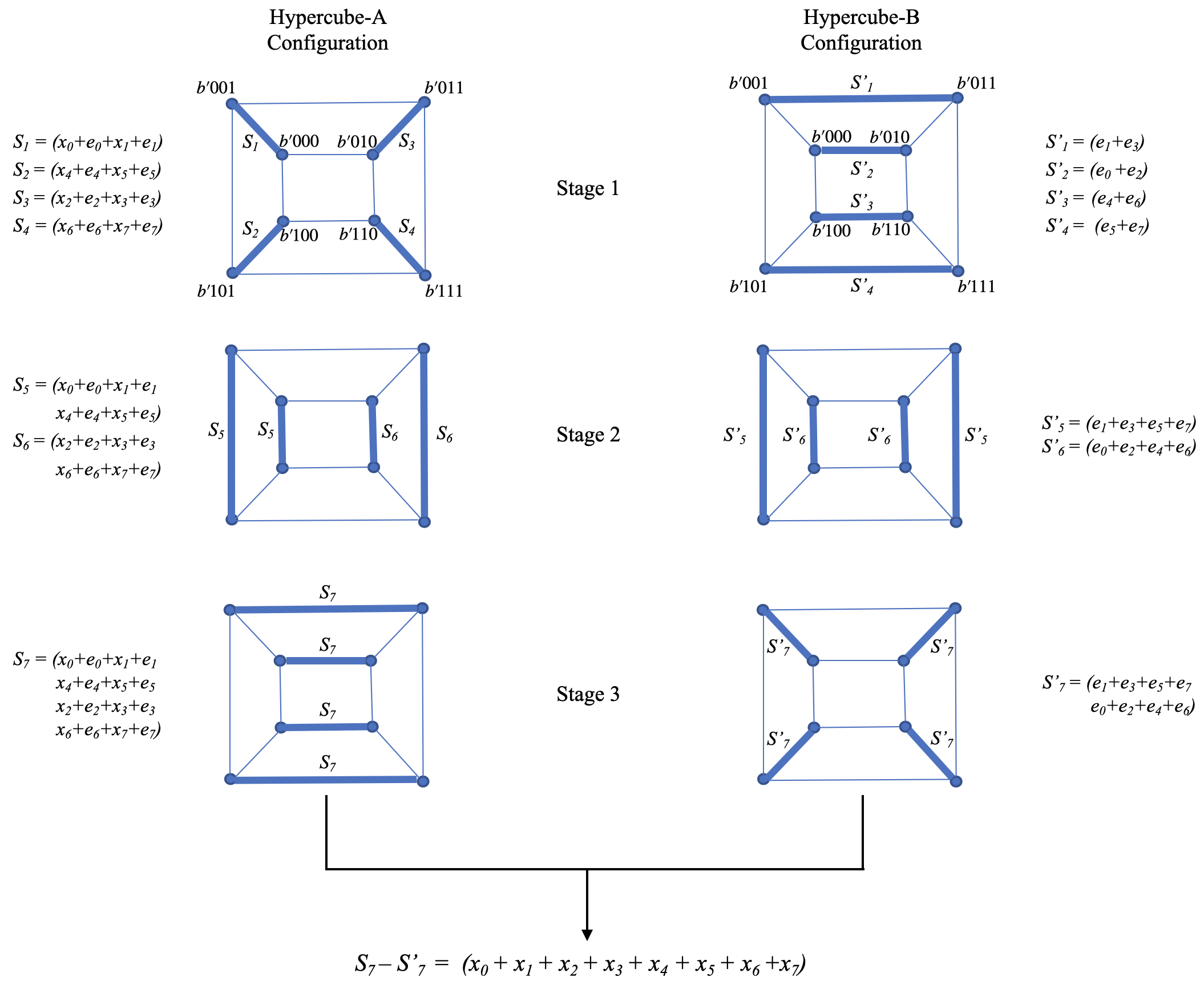}}
\caption{An Example of Protocol 2 Calculations on Two 3-dimensional (8-node) Hypercubes}
\end{figure*}

\subsection{Zero-knowledge Proof Model}

In the proposed private data aggregation protocol, the straightforward public summations of values are replaced with the private summations validated through zero-knowledge proof scheme. The pseudocode of this zero-knowledge proof scheme is presented in Algorithm A.1 where there are two different data types as private and public. While private data is only known by a party, public data is publicly available in the smart contract. In line 2, the hash of the private summation of the first party is computed along with the salt value to mask. In line 3, the hash of the private summation of the second party paired with the first party at that stage of the hypercube network is also computed. In line 4, the resulting private summation is computed with the new salt value. Later, Zokrates checks three different conditions: (i) the correctness of the commitment of the first party, (ii) the correctness of the commitment of the second party and finally (iii) the correctness of the commitment after the summation of the private values. The parties cannot cheat by providing wrong private values since the commitments are already stored in the smart contract.

There exist several important points in this structure that should be noted. Firstly, suppose a scenario where no salt value is used to emphasize the importance of salt values. Then, a malicious party may easily perform a preimage attack by using an enumeration table to resolve the commitments. The salt value of a party is replaced with a novel value at each time that party generates a proof to increase the security level further. Secondly, although the private summations of two parties paired result in the same private value after the interaction at that stage of the hypercube network, their commitments in the smart contract are totally different because of the different salt values. Thirdly, while each zero-knowledge proof is generated off-chain using the web user interface application, it is submitted and verified on-chain in the smart contract. If it is correctly verified, the commitment is updated with the new commitment in the contract, indicating the summation performed. Lastly, the binding property of the SHA-256 scheme prevents the association of the commitment with another value later while its hiding property prevents the extraction of any useful information from the commitment.

\subsection{Smart Contract Model}

The smart contract components of the proposed protocol provide required functions to a group of parties to perform decentralized and privacy-preserving data aggregation. These functions include \textit{user registration}, \textit{private summation submission} and \textit{proof verification}. The first function is \textit{user registration} where a party registers to a private summation group through the commitments of the secret and masking values; and a public key. Once the party registers, it cannot register any more to prevent a Sybil attack where a malicious party would perform more than one registration to disrupt the normal flow of the protocol and to learn the secret values of the other parties. The pseudo-code of the functions is given in Algorithm A.2.

The second function of the protocol is \textit{private summation submission} where a party is systematically paired with two other parties in the hypercube network at each stage and submits a private summation through a direct communication channel. This function takes the cipher-texts as input, which are encrypted with asymmetric encryption beforehand. The pairings are identified on-chain with exclusive-or (i.e. XOR) operation and therefore, the smart contract must track the current stage of the hypercube network. Once a party submits the encrypted messages to the peer parties, the corresponding positions in the two-dimensional mappings are updated. The pseudo-code of the function is given in Algorithm A.3.

The third function is \textit{proof verification} where a party verifies the correctness of zero-knowledge proof on-chain. The function takes the proofs and the next commitments calculated after the private summation has been done off-chain. The proofs are verified if and only if the current commitments of the  party and the peer parties are correct and the next commitments of the party are correct as well. Therefore, the smart contract must necessarily track the commitments of the secret values of the parties. Otherwise, the parties would generate zero-knowledge proof for any values they wish. The verification of proofs are completed after calling the function \textit{verifyTx} from an external smart contract which is generated by the Zokrates toolbox and deployed to the blockchain. If the proofs are correct, the number of proofs is incremented so that the current stage of the hypercube network is incremented as well whenever the number of proofs is equal to the number of registered users. Furthermore, the concept of temporary commitment must be applied to change the commitments of the paired parties at the same time. Otherwise, the proof of the party submitting it to the smart contract later would not be able to be correctly verified. The pseudo-code of the function is given in Algorithm A.4. 

\subsection{Web User Interface Model}

The interactions with the smart contract by considering the complexity of zero-knowledge proof, cryptographic commitment and asymmetric encryption operations would be very difficult for a party. Therefore, we have designed and implemented a web user interface application running in browsers without any trusted setup as long as the Metamask extension \cite{metamask} is already downloaded. In this application, there are four main functionalities as (i) \textit{contract deployment} where a party can easily deploy an instance of the contract, (ii) \textit{register} where a party can register to a private summation group, (iii) \textit{submit} where a party is paired with another party with respect to the stage of the hypercube network and submits private summation using direct private channel using the contract, and (iv) \textit{verify} where a party generates zero-knowledge proof to prove the correctness of the summation performed. 

The screenshot for the user interface of the first functionality, \textit{contract deployment}, is provided in Fig. 6. where it requires two values as registration start time and the registration time limit in terms of block time to define the start and the finish time of the registration period. After the registration period finishes, no other party can join the group further. The registration period is also important to compute the number of necessary stages for the hypercube network. The more parties register to a private summation group, the more number of stages is logarithmically required. Once the button is clicked, the application communicates with Ethereum to deploy a new contract instance and returns the contract address. The owner of the contract can share that address to form a private summation group.

\begin{figure*}[htbp]
\centerline{\includegraphics[width=0.7\textwidth]{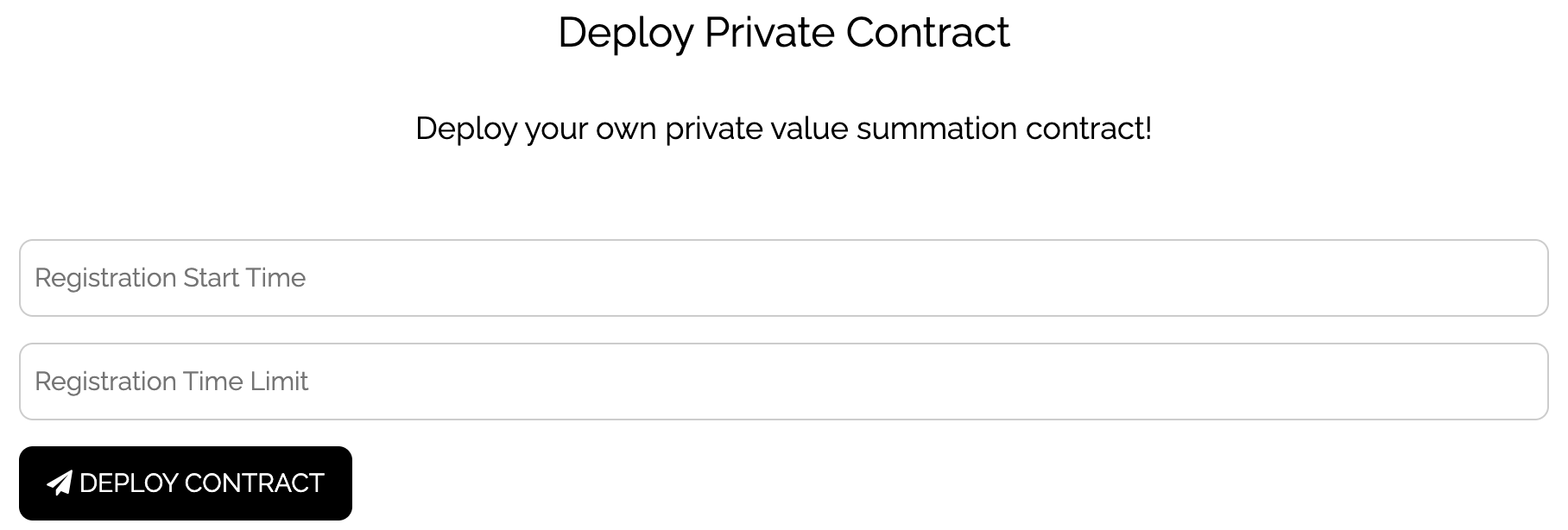}}
\caption{Web User Interface for Contract Deployment}
\end{figure*}

The screenshot for the user interface of the second functionality, \textit{user registration}, is provided in Fig. 7. where it requires the contract address and the secret value in order to compute its commitment. The public key of the registering party is also automatically restored so that it can be used in later phases for asymmetric encryption. The application submits the commitment of a random value and the commitment of a masked value of the secret value with that random value. Later, it returns the transaction address and a secure number. In the protocol, this secure number is a tuple with four hex values as (i) the masked value of the secret value, (ii) the random value, (iii) the salt value for the masked value and (iv) the salt value for the random value, respectively as $(h'(x+e), h'(e), h'(r_{x}), h'(r_{e}))$.

\begin{figure*}[htbp]
\centerline{\includegraphics[width=0.7\textwidth]{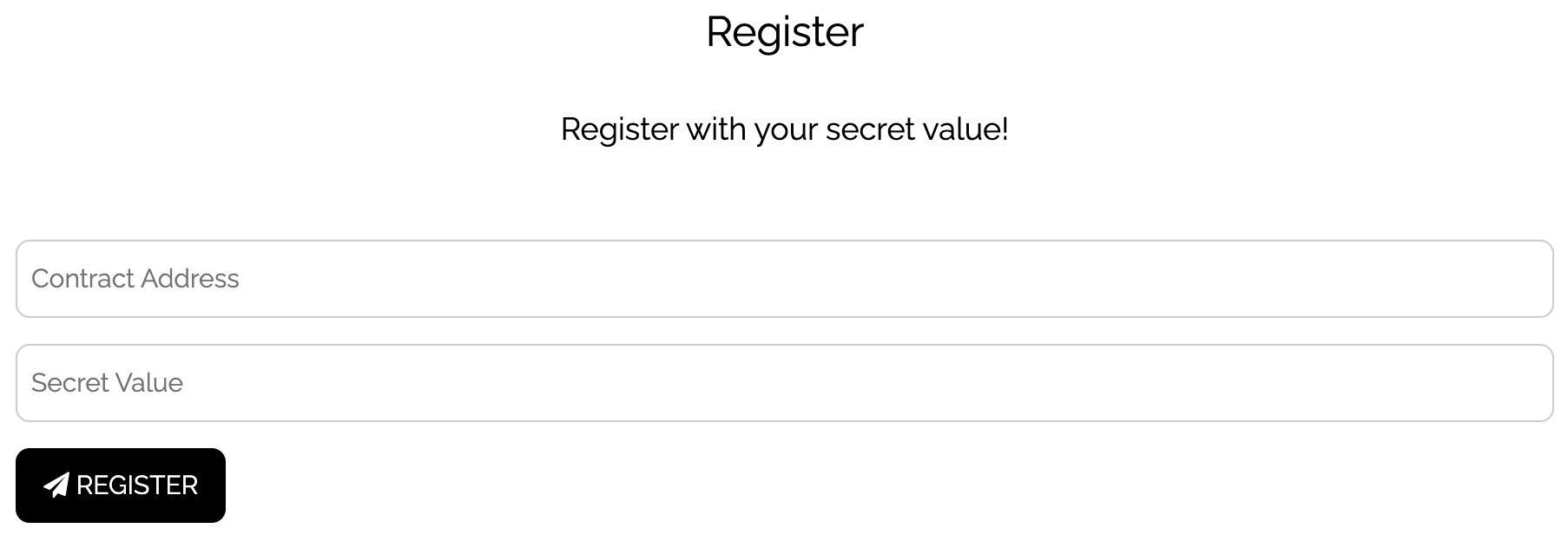}}
\caption{Web User Interface for User Registration}
\end{figure*}

The screenshot for the user interface of the third functionality, \textit{private summation submission}, is provided in Fig. 8. where it requires the contract address and the secure number mentioned previously. Once the button is clicked, the application fetches the public keys of the parties paired at that stage of the hypercube network from the smart contract in order to encrypt the private summation. It submits the encrypted messages to the parties paired through the smart contract and returns the resulting transaction address.

\begin{figure*}[htbp]
\centerline{\includegraphics[width=0.7\textwidth]{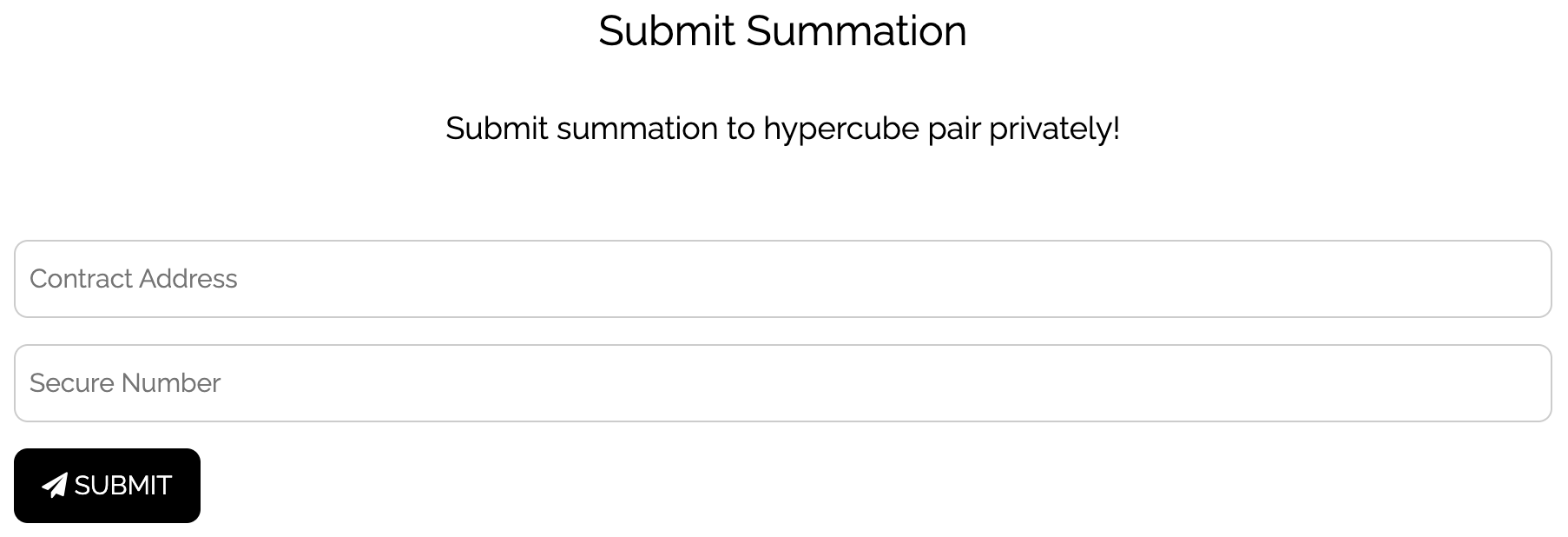}}
\caption{Web User Interface for Private Summation Submission}
\end{figure*}

The screenshot for the user interface of the fourth and the final functionality, \textit{proof verification}, is provided in Fig. 9. where it requires the contract address, the secure number and the private key of the party which is used only for the browser operations. Once the button is clicked, the application fetches the secret messages of the parties paired from the smart contract and decrypts them to resolve the private summations of the parties paired. The private summations are summed and the zero-knowledge proofs about the correctness of these summations are generated. Later, the proofs are submitted and verified in the smart contract.

\begin{figure*}[htbp]
\centerline{\includegraphics[width=0.7\textwidth]{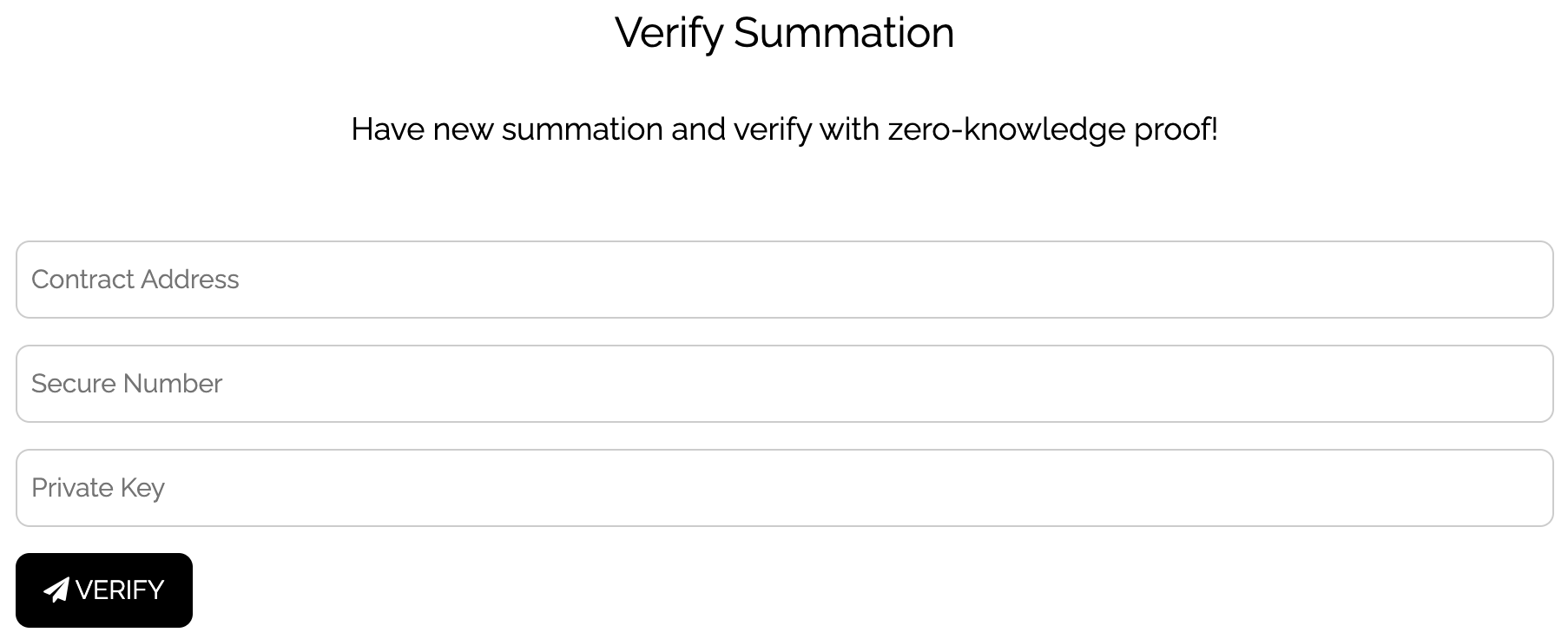}}
\caption{Web User Interface for Proof Verification}
\end{figure*}

\section{Analysis of Protocol}

The proposed protocol is formally analyzed with respect to the security and scalability perspectives. While the security analysis focuses on the correctness of the schemes used in the protocol as well as the vulnerability attacks, the scalability analysis focuses more on the computational, communicational and storage overheads. The extension of the protocol to the incomplete hypercube is also discussed for future work. 

\subsection{Security}

The security of the protocol in terms of the underlying approaches used in the protocol and the vulnerability attacks is analysed with the following way:

\subsubsection{Zero-Knowledge Proof Security} The underlying zk-SNARKs approach implemented through the Zokrates framework is proven to be secure in the literature \cite{sasson2014, eberhardt2018}.

\subsubsection{Encryption Security} The ECIES encryption and decryption function used during the communications in the hypercube networks is proven to be secure in the literature \cite{martinez2015}.

\subsubsection{Commitment Security} The SHA-256 commitment scheme is also proven to be secure in the literature \cite{sha256, gilbert2003}.

\subsubsection{Preimage Attack to Commitment Security} The preimage attack is a cryptographic attack used to break the commitment functions and consequently the commitments themselves. In the preimage attack, an enumeration (i.e. look-up) table is utilized for the comparison of the commitments with the brute-force approach until the correct comparison is found. In this work, each secret value is masked through a salting value generated by the pseudo-random number generator $Prng$:

\begin{align}
& c = Comm(x, Prng()) 
\end{align}

\noindent
where $c$ is the resulting commitment of the commitment function $Comm$ based upon the secret value $x$ and the random number generator function $Prng()$ which makes selection from $2^{53}$ possible values. In the SHA-256 commitment scheme, even a minor change in the inputs leads to a significant change in the resulting output due to the avalanche effect \cite{sha256}. In the smart contract of our protocol, the commitments stored would be totally different even though the secret values of two distinct parties are the same. Furthermore, the corresponding salting value has been changing every time a new zero-knowledge proof is generated.  

\subsubsection{Data Aggregation Security} The proposed protocol utilizes two hypercube networks where the first network is responsible for aggregating the secret values masked through the addition with the random values while the second network is responsible for aggregating the random values themselves:

\begin{align}
& c^A = Comm(x + Prng(), Prng()) \\
& c^B = Comm(Prng(), Prng())
\end{align}

\noindent
where $c^A$ and $c^B$ are the resulting commitments of the commitment function $Comm$, based on the secret value $x$, the random value and the salting values generated by pseudo-random number generator $Prng$ for \textit{Configuration-A} and \textit{Configuration-B}, respectively. The addition of the random values to the secret values naturally leads the results to look random as well. After all the aggregations in the hypercube networks, the summation can be securely found as $\sum (x) = \sum (x + Prng()) - \sum (Prng())$.

\subsubsection{Collusion Attack} In the protocol, the users learn no information at all about the secret values of the other users. The secret value of a specific user could be revealed if and only if the users that communicate with that user in the first stages of two hypercube configurations collaborate with each other. In the most generalized version, the more hypercube configurations are utilized, the more secure the system will be in return of communicational, computational and storage overheads. However, at least two hypercube configurations are mandatory to provide the fully-secure privacy-preserving data aggregation protocol.

\subsection{Scalability}

The scalability of the protocol with an increasing number of parties in terms of the communicational, computational and storage overheads is analysed with the following way:

\subsubsection{Communicational Overhead} The main factor affecting the communication overhead is the inherent characteristics of the hypercube network topology. Each user need to perform $2 \times log(N)$ number secret message exchanges with their peers, resulting in $2 \times N \times log(N)$ number of exchanges at total for all users. The overhead of running two hypercube networks is alleviated through parallelism in the web user interface application without any usability effect.

\subsubsection{Computational Overhead} The main computational bottleneck is the off-line zero-knowledge proof generations. For the protocol, a user involved with a group to perform privacy-preserving data summation has to generate two distinct proofs for every stage of the hypercube network topology. For a group with $N$ number of users, $2 \times log(N)$ number of proofs must be generated for each individual user, resulting in $2 \times N \times log(N)$ number of proofs in total for all users. 

\subsubsection{Storage Overhead} The storage overhead of the protocol is originated from the smart contracts running in the blockchain. For the protocol, eight different mappings have been used to store (i) one for the registrations, (ii) two for the commitments in two hypercube networks, (iii) two for the temporary commitments, (iv) one for the public keys and (v) two for the secret messages between users.

The protocol performance on these overheads with an increasing number of parties is given in Table B.1 for each party and for the whole system. The communicational, computational and storage overheads are presented through the number of proofs generated, the number of exchanges performed and the number of keys in the mappings created, respectively. The overheads for each party is more important to evaluate the protocol since the system functions are completely distributed over all the users. However, the overheads for the whole system is used to detect the system throughput in terms of the block interval since the transactions of all users have to be written to blockchain. Overall, it could be said that the protocol provides a secure way to perform privacy-preserving data aggregation in return of certain communicational, computational and storage overheads. It is appropriate to be used in scalable systems with a high number of users with independent functionalities distributed over all the users. 

\subsection{Extension to Incomplete Hypercube}

The number of parties running the privacy-preserving data aggregation protocol may not satisfy the following condition at all times, $N = 2^{d}$ where $d$ is the total number of dimensions in the hypercube network topology. This leads to the concept of incomplete hypercube where not every party may necessarily have a peer party at every hypercube communication stage and some links are missing \cite{katseff1988}. There exist techniques proposed in the literature to provide communications between parties in such cases. According to the work \cite{prabhala1992}, any incomplete hypercube with number of parties $N = 2^{d} + N^{'}$ can be considered as the collection of several successively smaller complete hypercubes as $N = 2^{d_1} + 2^{d_2} + ... + 2^{d_i}$ where $d_1 > d_2 > ... > d_i$ and $i$ is the number of complete hypercubes required. We will investigate applying this approach to extend our protocol for an arbitrary number of parties as a future work.

\section{Experimental Study}

In the experimental study, the proposed protocol is exposed to several tests to measure (i) the individual and the total gas costs on the Avalanche Fuji blockchain and the local Ethereum blockchain and (ii) the zero-knowledge proof generation and verification times. In order to increase the reproducibility of our tests and to validate the correctness of our protocol, the parameters settings and the source code of the system are publicly shared.

\subsection{Experimental Setup}

The proposed private data aggregation protocols consist of two main components as the smart contracts in the Ethereum blockchain and the web user interface application running in the browser. For the implementation of the smart contracts, the Solidity language \cite{solidity} and the 0.8.0-version of Remix compiler \cite{remix} are used. For the implementation of the frontend and the backend designs of the web user interface application, HTML, CSS and Javascript languages are used, respectively. The following libraries are also used in the backend implementation: (i) the \textit{SHA-256} commitment scheme \cite{sha256} due to the built-in integration with the Zokrates toolbox: (ii) the \textit{eccrypto} javascript library \cite{eccrpyto} to encrypt and to decrypt secret messages in the direct private channels between parties, (iii) the 0.8.3-version of the \textit{Zokrates} toolbox \cite{eberhardt2018} and the 1.1.4-version of the \textit{zokrates-js} javascript library \cite{zokratesjs} to generate and to verify zero-knowledge proofs, (iv) the \textit{ethers} library \cite{ethers} to bridge the browser to the Ethereum blockchain with Metamask \cite{metamask}. The browserify \cite{browserify} and webpack \cite{webpack} bundlers are chosen to integrate these libraries into the browser. 

In the experimental study, the performance of the protocols has been evaluated with respect to the following measures: (i) verifications of the system requirements, (ii) the Ethereum gas cost consumed and (iii) the zero-knowledge proof generation and verification times. All the experimental tests are performed on MacBook Pro Notebook with 2.6 GHz Intel Core i7 processor, 16 GB memory and 6 cores. The ready-to-use implementations of the proposed private data aggregation protocols are made available in our GitHub page \cite{github}.

\subsection{Requirement Verification}

In this section of the paper, the protocol is evaluated with respect to the satisfactions of the problem requirements that are previously defined. The overall results of the requirement verification are summarized in Table 2 for the requirements including privacy, confidentiality, public verifiability, authentication, non-interactivity, scalability, correctness and usability; and for the system components including the Ethereum blockchain, smart contract, commitment scheme, asymmetric encryption, zero-knowledge proof scheme, hypercube network topology and web user interface application. 

According to Table 2, for privacy, zero-knowledge proofs based on the commitments of secret values are generated and verified in the protocol, (\textit{R1}). For confidentiality, asymmetric encryption between users is necessary to protect the data from the external environment, (\textit{R2}). For trustless, the pairwise communication guaranteed through the cryptographic primitives between parties in hypercube network topology eliminates the need for any trusted third party, (\textit{R3}). For public verifiability, Ethereum and the smart contracts provides an underlying platform and functions in order to verify the correctness of zero-knowledge proofs, (\textit{R4}). For authentication, the registration period defined in the smart contracts prevents unauthenticated users to involve with the data summation in later periods, (\textit{R5}). For non-interactivity, zk-SNARKs-based zero-knowledge proof scheme does not require any form of interactivity between prover and verifier to generate and verify the proof, (\textit{R6}). However, note that the interactions between parties in terms of the overall system are still required to exchange certain encrypted messages to keep the protocol running. For scalability, the underlying blockchain infrastructure and the hypercube network topology lead a very high number of users to use the system in return for logarithmic communicational overhead, (\textit{R7}). For correctness, the components except for hypercube network topology and web user interface are necessary. As an example, the protocol is tailored for Ethereum and they may not necessarily be correct in another blockchain, (\textit{R8}). For usability from the perspective of individual users, web user interface plays an important role, (\textit{R9}). 

\begin{table}[htbp]
\caption{Verification of Problem Requirements}
\footnotesize
\begin{center}
\resizebox{\columnwidth}{!}{%
\begin{tabular}{|l|l|l|l|l|l|l|l|}
\hline
\textbf{Requirement} & \rotatebox{90}{\textbf{Ethereum Blockchain}} & \rotatebox{90}{\textbf{Smart Contract}} & \rotatebox{90}{\textbf{Commitment Scheme}} & \rotatebox{90}{\textbf{Asymmetric Encryption}} & \rotatebox{90}{\textbf{Zero-Knowledge Proof}} & \rotatebox{90}{\textbf{Hypercube Topology}} & \rotatebox{90}{\textbf{Web User Interface}} \\
\hline
R1. Privacy & & & \checkmark & & \checkmark & & \\
\hline
R2. Confidentiality & & & & \checkmark & & & \\
\hline
R3. Trustless & & & \checkmark & \checkmark & \checkmark & \checkmark & \\
\hline
R4. Public Verifiability & \checkmark & \checkmark & & & \checkmark & & \\
\hline
R5. Authentication & & \checkmark & & & & & \\
\hline
R6. Non-interactivity & & & & & \checkmark & & \\
\hline
R7. Scalability & \checkmark & & & & & \checkmark & \\
\hline
R8. Correctness & \checkmark & \checkmark & \checkmark & \checkmark & \checkmark & & \\
\hline
R9. Usability & & & & & & & \checkmark  \\
\hline
\end{tabular}
}
\end{center}
\footnotesize
\end{table}

\subsection{Ethereum Gas Costs}

The gas consumption of the functions in the smart contracts are measured using the Avalanche Fuji testnet \cite{fuji} and the local Ethereum blockchain with varying number of parties. The results of the measurements are presented in Table 3 in terms of gas units, gas cost in Avax and gas cost in USD. The gas unit for each distinct function is the average of the gas costs of all the transactions performed by the parties. Accordingly, the higher number of parties leads to more transactions for a certain function to be accumulated. The maximum number of parties tested is 128 in our experiments where it takes nearly 24 hours. However, the amount of time needed for the completion of the proposed protocol is considerably less in the real life scenarios due to the parallel computation of the parties.

The functions include contract deployment, user registration, private summation submission and finally proof verification. As of 30/12/2022, the base gas price per gas unit is 25 nAvax for Avalanche Fuji testnet. According to Table 3, the most expensive function is the contract deployment with approximately 3 million gas units, which corresponds to more than 2 \$. The main reason behind this high cost is the function call to an external smart contract to verify zero-knowledge proofs on-chain. However, the least expensive function is the user registration since it affects only certain mappings in the contracts.

\begin{table*}[htbp]
\caption{The gas costs of the functions in the smart contract}
\scriptsize
\begin{center}
\begin{tabular}{|l|r|r|r|r|r|r|r|}
\hline
\cellcolor{gray!50} \textbf{Function} & \cellcolor{gray!50} \makecell[l]{\textbf{Gas Units} \\ \textbf{(Fuji)}}  & \cellcolor{gray!50} \makecell[l]{\textbf{Gas Units} \\ \textbf{(Local 2)}} & \cellcolor{gray!50} \makecell[l]{\textbf{Gas Units} \\ \textbf{(Local 8)}} & \cellcolor{gray!50} \makecell[l]{\textbf{Gas Units} \\ \textbf{(Local 32)}} & \cellcolor{gray!50} \makecell[l]{\textbf{Gas Units} \\ \textbf{(Local 128)}} & \cellcolor{gray!50} \makecell[l]{\textbf{Gas Cost} \\ \textbf{(Avax)}} & \cellcolor{gray!50} \makecell[l]{\textbf{Gas Cost} \\ \textbf{(USD)}} \\
\hline
\makecell[l]{\textbf{Contract Deployment}} & 3,372,418  & 3,412,242 & 3,412,242 & 3,412,242 & 3,412,242 &  0.0927415 & 2.12 \$ \\
\hline
\makecell[l]{\textbf{User Registration}} & 253,118 & 264,654 & 259,025 & 257,014 & 256,360 & 0.0075556 & 0.17 \$ \\
\hline
\makecell[l]{\textbf{Private Summation Submission}} & 763,692 & 761,509 & 763,086 & 763,401 & 763,537 & 0.0205294 & 0.47 \$ \\
\hline
\makecell[l]{\textbf{Proof Verification}} & 2,030,995 & 2,010,273 & 1,998,132 & 1,996,382 & 1,996,063 & 0.0547150 & 1.25 \$  \\
\hline
\end{tabular}
\end{center}
\scriptsize
\end{table*}

With an increasing number of parties (from 2 to 128 parties) in the local blockchain, while the gas consumption of the contract deployments are unaffected, there exists a slight decrease in the gas consumption of the user registration. It has been observed that the initial transactions for that function tend to use more gas units while the following transactions neutralize this effect due to the averaging operation. The more transactions there are with an increasing number of parties involved, the more neutralized it eventually becomes. The similar phenomenon is also observed for the proof verification where the certain transactions at the end of each hypercube dimension use more gas units (Line 10 of Algorithm 4). Assuming $N$ parties with $log(N)$ dimensions, there are $log(N)$ transactions with extra gas units ($tx^+$) and $N - log(N)$ transactions with normal gas units ($tx$). Based on the fact $tx < tx^+$, the average of the transactions is:

\begin{equation}
\frac{log(N) \cdot tx^+ + (N - log(N)) \cdot tx}{N}
\end{equation}
\noindent 
After simplifying, we get:
\begin{equation}
tx + \frac{log(N)}{N} \cdot (tx^+ - tx)
\end{equation}
\noindent
Note that $(tx^+ - tx)$ is always positive and in the limit, 
$\lim_{N\to\infty} \frac{log(N)}{N} = 0. $

The gas costs for high number of parties and correspondingly, the amount of gas costs can be generalized as follows for each individual user:

\begin{align}
& GC_{total} = GC_{fixed} + log(N) \times GC_{dynamic}  \nonumber \\
& GC_{fixed} = GC_{registration} \nonumber \\
& GC_{dynamic} = GC_{submission} + GC_{verification}  \nonumber \\
\end{align}

\noindent
where $GC_{registration}, GC_{submission}$ and $GC_{verification} $ denote the gas costs for the user registration, private summation submission and the proof verification functions, respectively. The total gas cost model for the second protocol with $N$ number of users is as follows: 

\begin{equation}
GC_{total} = N \times GC_{fixed} + 2 \times N \times log(N) \times GC_{dynamic}
\end{equation}

\subsection{Zero-Knowledge Proof Generation/Verification Times}

The  zero-knowledge proof generation times have been calculated through 20 independent runs in the browser. The corresponding results are presented in Fig. 10 where each dot represents an individual measurement in seconds while the box plot represents overall statistics about these measurements. With respect to the given figure, the average proof generation time is approximately 88.3 seconds while the median time is 87.4 seconds. The maximum proof generation time is 95.8 seconds where it belongs to the first experimental run due to the cold start effect. Note that the box plot marks it as an outlier since it resides over the upper fence. On the other hand, the minimum proof generation time is 87.1 seconds. As understood from the given statistics, the distribution of the proof generation times are uneven and positively skewed. The complete data for the experiment is given in Table B.2. On the other hand, since the zero-knowledge proofs are verified on-chain in the smart contracts, they do not have any temporal costs other than the gas costs.

\begin{figure}[htbp]
\centerline{\includegraphics[width=0.25\textwidth]{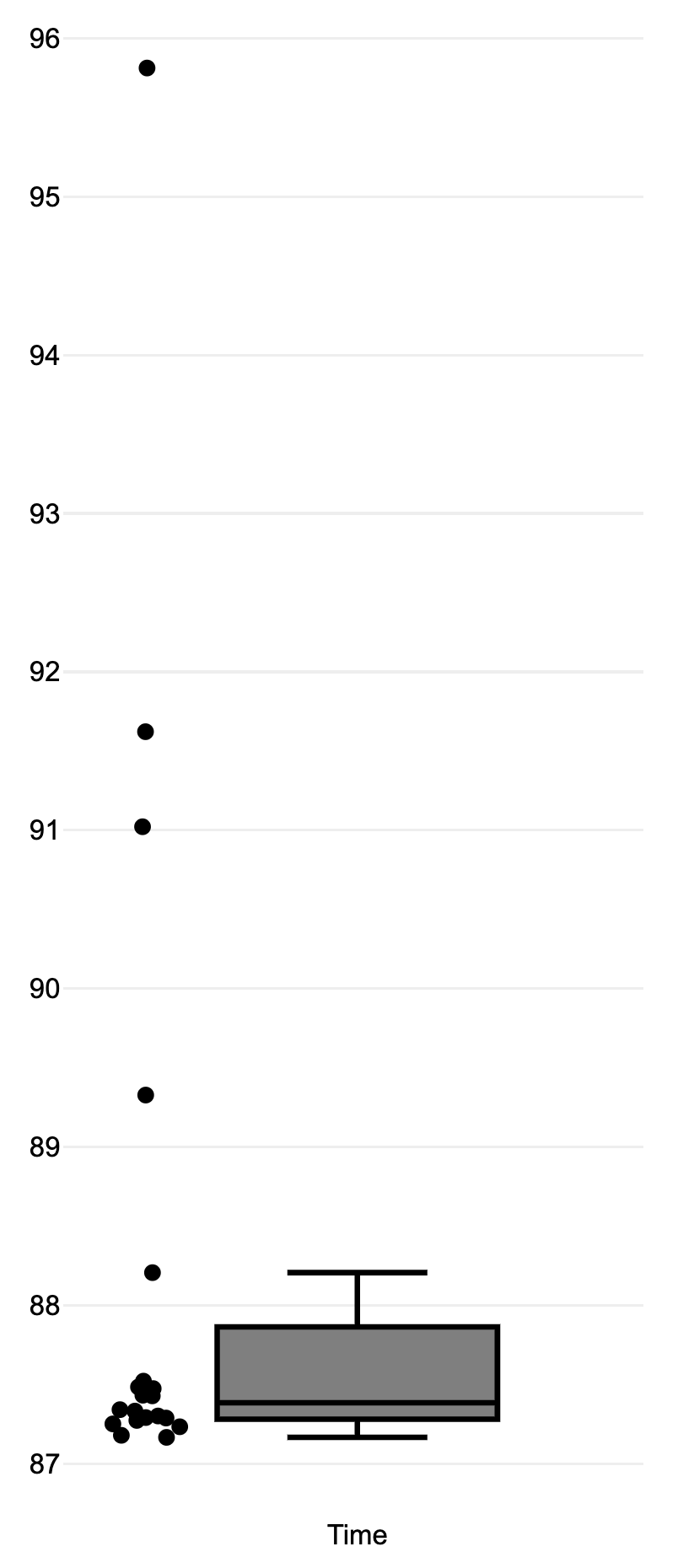}}
\caption{Box Plot for Zero-Knowledge Proof Generation Times}
\end{figure}

\section{Conclusions and Future Work}

In this paper, we have proposed a privacy-preserving data aggregation protocol for data summation with the help of the cryptographic operations and the hypercube network on the Ethereum blockchain. In the first phase, the parties register and form a private summation group after an owner party deploys an instance contract. In the second phase, the parties paired to communicate with each other through the direct private channel by means of the smart contract. In the third phase, they perform the summation operations and generate the zero-knowledge proof in order to prove the correctness of the summation operations. In the paper, the details for the smart contracts, the zero-knowledge proof scheme and the web user interface design are also presented. The proposed protocol has been analyzed from two main standpoints as security and scalability. In the experimental study, we have identified the gas costs required per party and per system. For a high number of users, the general formulation of the changes in the gas cost is also presented. 

In the future, it is planned to reduce the communication overhead between different parties in the network and the zero-knowledge proof generation times to a certain extent. The proposed protocol might be extended to support an arbitrary number of parties. Furthermore, it is also planned to compare other network topologies including shuffle-exchange and De Bruijn networks with the currently used hypercube network.

{\appendices

\section{Algorithms}

The algorithms used in our work are presented for (i) the private summation proof in Zokrates in Algorithm A.1, (ii) the user registration in the smart contract in Algorithm A.2, (iii) the private summation submission in the smart contract in Algorithm A.3; and finally (iv) the summation proof verification in the smart contract in Algorithm A.4.

\begin{algorithm*} 
\renewcommand\thealgorithm{A.1}
\small
\caption{Zokrates Code for Private Summation}  
\begin{algorithmic}[1]

\STATE \textbf{def} main(\textbf{private} sum, \textbf{private} salt, \textbf{private} saltNext, \textbf{private} sumPair, \textbf{private} saltPair, \textbf{public} sumHash, \textbf{public} sumPairHash, \textbf{public} sumNextHash):

\STATE \;\;\;\; computedSumHash \textbf{=} \textbf{sha256}(sum, salt)

\STATE \;\;\;\; computedSumPairHash \textbf{=} \textbf{sha256}(sumPair, saltPair)

\STATE \;\;\;\; computedSumNextHash \textbf{=} \textbf{sha256}(sum \textbf{+} sumPair, saltNext)

\STATE \;\;\;\; result = \textbf{if}(computedSumHash \textbf{==} sumHash \textbf{\&\&} computedSumPairHash \textbf{==} sumPairHash \textbf{\&\&} computedSumNextHash \textbf{==} sumNextHash) \textbf{then} 1 \textbf{else} 0 \textbf{fi}

\STATE \;\;\;\; \textbf{return} result 

\end{algorithmic} 
\small
\end{algorithm*}

\begin{algorithm*} 
\renewcommand\thealgorithm{A.2}
\small
\caption{User Registration}  
\begin{algorithmic}[1]

\STATE \textbf{function} register(\textbf{uint} commitmentA, \textbf{uint} commitmentB, \textbf{string} publicKey) \{

\STATE \;\;\;\; \textbf{require}(\textbf{bytes}(publicKeys[msg.sender]).length \textbf{==} 0);

\STATE \;\;\;\; registrationIds[msg.sender] \textbf{=} registeredUsers.length;

\STATE \;\;\;\; commitmentsA[msg.sender] \textbf{=} commitmentA;

\STATE \;\;\;\; commitmentsB[msg.sender] \textbf{=} commitmentB;

\STATE \;\;\;\; publicKeys[msg.sender] \textbf{=} publicKey;

\STATE \;\;\;\; registeredUsers.\textbf{push}(msg.sender);

\STATE \}

\end{algorithmic} 
\small
\end{algorithm*}

\begin{algorithm*} 
\renewcommand\thealgorithm{A.3}
\small
\caption{Private Summation Submission}  
\begin{algorithmic}[1]

\STATE \textbf{function} submit(\textbf{string} encryptedMessageA, \textbf{string} encryptedMessageB) \{

\STATE \;\;\;\; \textbf{uint256} hypercubePairA \textbf{=} registrationIds[msg.sender] \textbf{\^} (2 \textbf{**} currentHypercubeStage);

\STATE \;\;\;\; \textbf{uint256} hypercubePairB \textbf{=} registrationIds[msg.sender]\textbf{ \^} (2 \textbf{**} (maximumHypercubeStage \textbf{-} currentHypercubeStage \textbf{-} 1));

\STATE \;\;\;\; \textbf{address} hypercubePairAddressA \textbf{=} registeredUsers[hypercubePairA]; 

\STATE \;\;\;\; \textbf{address} hypercubePairAddressB \textbf{=} registeredUsers[hypercubePairB]; 

\STATE \;\;\;\; secretMessagesA[msg.sender][hypercubePairAddressA] \textbf{=} encryptedMessageA;

\STATE \;\;\;\; secretMessagesB[msg.sender][hypercubePairAddressB] \textbf{=} encryptedMessageB;

\STATE \}

\end{algorithmic} 
\small
\end{algorithm*}

\begin{algorithm*} 
\renewcommand\thealgorithm{A.4}
\small
\caption{Proof Verification}  
\begin{algorithmic}[1]

\STATE \textbf{function} verify(\textbf{Proof} proofA, \textbf{Proof} proofB, \textbf{uint} nextCommitmentA, \textbf{uint} nextCommitmentB) \{

\STATE \;\;\;\; \textbf{uint256} hypercubePairA \textbf{=} registrationIds[msg.sender] \textbf{\^} (2 \textbf{**} currentHypercubeStage);

\STATE \;\;\;\; \textbf{uint256} hypercubePairB \textbf{=} registrationIds[msg.sender]\textbf{ \^} (2 \textbf{**} (maximumHypercubeStage \textbf{-} currentHypercubeStage \textbf{-} 1));

\STATE \;\;\;\; \textbf{address} hypercubePairAddressA \textbf{=} registeredUsers[hypercubePairA]; 

\STATE \;\;\;\; \textbf{address} hypercubePairAddressB \textbf{=} registeredUsers[hypercubePairB]; 

\STATE \;\;\;\; \textbf{bool} isProofCorrectA \textbf{=} verifier.verifyTx(proofA, [commitmentsA[msg.sender], commitmentsA[hypercubePairAddressA], nextCommitmentA, 1]);

\STATE \;\;\;\; \textbf{bool} isProofCorrectB \textbf{=} verifier.verifyTx(proofB, [commitmentsA[msg.sender], commitmentsB[hypercubePairAddressB], nextCommitmentB, 1]);

\STATE \;\;\;\; \textbf{if} (isProofCorrectA \textbf{\&\&} isProofCorrectB)  \{

\STATE \;\;\;\;\;\;\;\; numberOfProofsVerified \textbf{+=} 1;

\STATE \;\;\;\;\;\;\;\; \textbf{if} (numberOfProofsVerified \textbf{==} registeredUsers.length)  \{

\STATE \;\;\;\;\;\;\;\;\;\;\;\; currentHypercubeStage \textbf{+=} 1;

\STATE \;\;\;\;\;\;\;\;\;\;\;\; numberOfProofsVerified \textbf{=} 0;

\STATE \;\;\;\;\;\;\;\; \}

\STATE \;\;\;\;\;\;\;\; \textbf{if} (tempCommitmentsA[hypercubePairAddressA] \textbf{==} 0)  \{

\STATE \;\;\;\;\;\;\;\;\;\;\;\; tempCommitmentsA[msg.sender] \textbf{=} nextCommitmentA;

\STATE \;\;\;\;\;\;\;\; \}

\STATE \;\;\;\;\;\;\;\; \textbf{else}  \{

\STATE \;\;\;\;\;\;\;\;\;\;\;\; commitmentsA[msg.sender] \textbf{=} nextCommitmentA;

\STATE \;\;\;\;\;\;\;\;\;\;\;\; commitmentsA[hypercubePairAddressA] \textbf{=} tempCommitmentsA[hypercubePairAddressA];

\STATE \;\;\;\;\;\;\;\;\;\;\;\; tempCommitmentsA[hypercubePairAddressA] \textbf{=} 0;

\STATE \;\;\;\;\;\;\;\; \}

\STATE \;\;\;\;\;\;\;\; \textbf{if} (tempCommitmentsB[hypercubePairAddressB] \textbf{==} 0)  \{

\STATE \;\;\;\;\;\;\;\;\;\;\;\; tempCommitmentsB[msg.sender] \textbf{=} nextCommitmentB;

\STATE \;\;\;\;\;\;\;\; \}

\STATE \;\;\;\;\;\;\;\; \textbf{else}  \{

\STATE \;\;\;\;\;\;\;\;\;\;\;\; commitmentsB[msg.sender] \textbf{=} nextCommitmentB;

\STATE \;\;\;\;\;\;\;\;\;\;\;\; commitmentsB[hypercubePairAddressB] \textbf{=} tempCommitmentsB[hypercubePairAddressB];

\STATE \;\;\;\;\;\;\;\;\;\;\;\; tempCommitmentsB[hypercubePairAddressB] \textbf{=} 0;

\STATE \;\;\;\;\;\;\;\; \}

\STATE \;\;\;\; \}

\STATE \}

\end{algorithmic} 
\small
\end{algorithm*}

\section{Protocol Performance}

The performance of the protocol is shown in Table B.1 in terms of the computation (as the number of proofs generated), communication (as the number of exchanges performed) and storage (as the number of keys in the mappings) overheads. The complete data of the zero-knowledge proof generation times for data aggregation is also presented in Table B.2.  

\begin{table*}[htbp]
\renewcommand\thetable{B.1}
\caption{Protocol Performance on Scalability Overheads}
\small
\begin{center}
  \begin{tabular}{|l|r|r|r|r|r|r|}
    \hline
      \cellcolor{black} &
      \multicolumn{2}{c}{\cellcolor{gray!50} \textbf{Computation Overhead (\#Proofs)}} &
      \multicolumn{2}{c}{\cellcolor{gray!50} \textbf{Communication Overhead (\#Exchanges)}} &
      \multicolumn{2}{c|}{\cellcolor{gray!50} \textbf{Storage Overhead (\#Keys)}} \\
    \hline
    \cellcolor{gray!50} \textbf{Number of Parties} & \cellcolor{gray!50} \textbf{Party} & \cellcolor{gray!50} \textbf{System} & \cellcolor{gray!50} \textbf{Party} & \cellcolor{gray!50} \textbf{System} & \cellcolor{gray!50} \textbf{Party} & \cellcolor{gray!50} \textbf{System} \\
    \hline
    2 & 2 & 4 & 2 & 4 & 8 & 16 \\
    \hline
    8 & 6 & 48 & 6 & 48 & 12 & 96 \\
    \hline
    32 & 10 & 320 & 10 & 320 & 16 & 512 \\
    \hline
    128 & 14 & 1792 & 14 & 1792 & 20 & 2560 \\
    \hline
  \end{tabular}
\end{center}
\small
\end{table*}

\begin{table}[htbp]
\renewcommand\thetable{B.2}
\caption{Zero-Knowledge Proof Generation Times}
\small
\begin{center}
\begin{tabular}{|c|c|}
\hline
\cellcolor{gray!50} \textbf{Run} & \cellcolor{gray!50} \textbf{Time (in sec)} \\
\hline
1 & 95.812 \\
\hline
2 & 91.022 \\
\hline
3 & 87.475 \\
\hline
4 & 91.622 \\
\hline
5 & 87.333 \\
\hline
6 & 88.207 \\
\hline
7 & 87.289 \\
\hline
8 & 87.485 \\
\hline
9 & 87.274 \\
\hline
10 & 87.433 \\
\hline
11 & 87.292 \\
\hline
12 & 89.328 \\
\hline
13 & 87.522 \\
\hline
14 & 87.167 \\
\hline
15 & 87.429 \\
\hline
16 & 87.342 \\
\hline
17 & 87.302 \\
\hline
18 & 87.234 \\
\hline
19 & 87.180 \\
\hline
20 & 87.252 \\
\hline
\end{tabular}
\end{center}
\small
\end{table}

}

\begin{IEEEbiography}[{\includegraphics[width=1in,height=1.25in,clip,keepaspectratio]{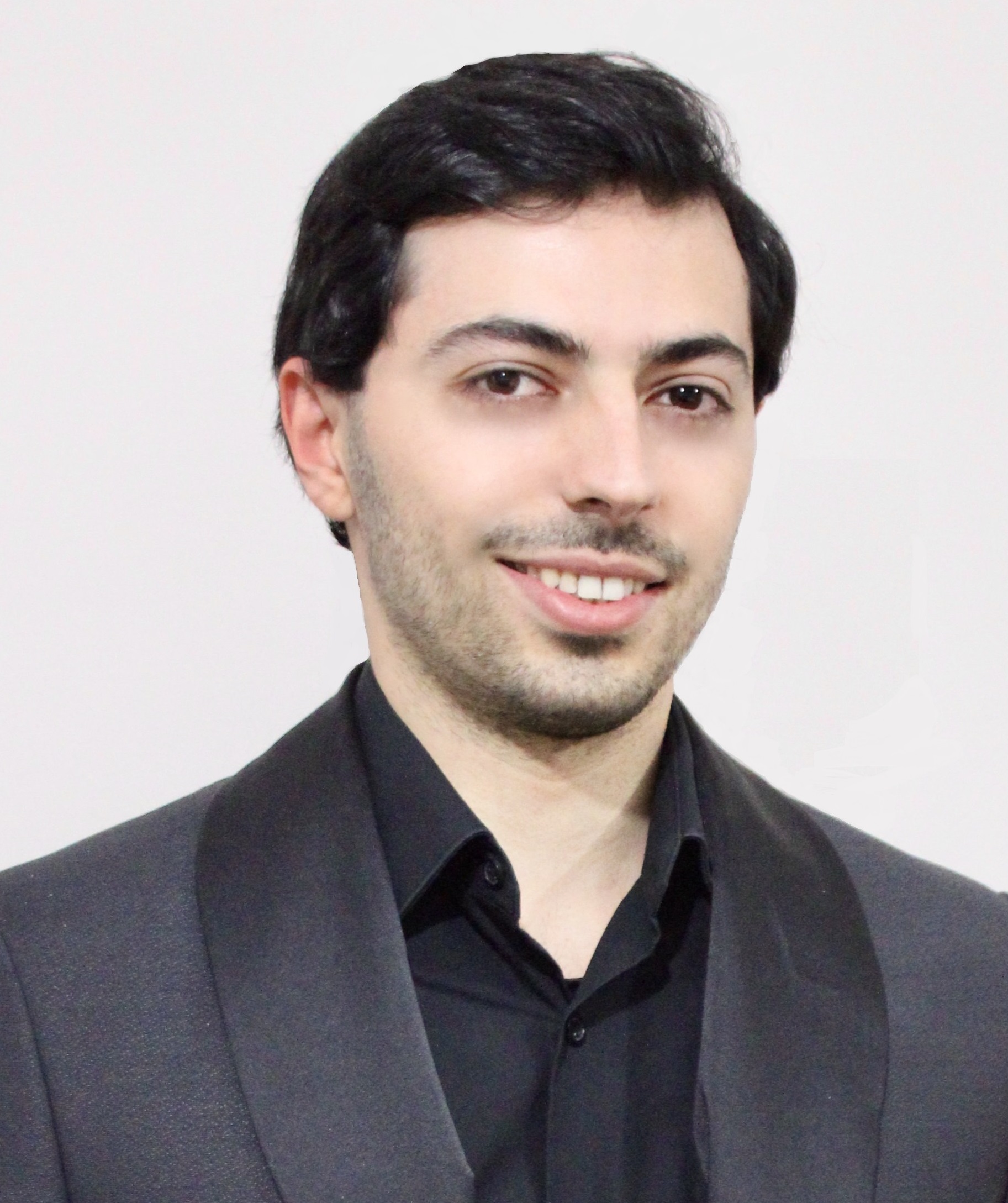}}]{Goshgar Ismayilov}
He is currently a Ph.D. candidate in Department of Computer Engineering at Bogazici Unversity. He received the B.S. degree and the M.S. degree in Department of Computer Engineering from Marmara University in 2016 and 2019, respectively. He was awarded to the first prize in the IEEE Computer Society Turkey Chapter Master Thesis in 2021. His main research interests include zero-knowledge proof, blockchain, dynamic optimization and cloud computing. 
\end{IEEEbiography}

\begin{IEEEbiography}[{\includegraphics[width=1in,height=1.25in,clip,keepaspectratio]{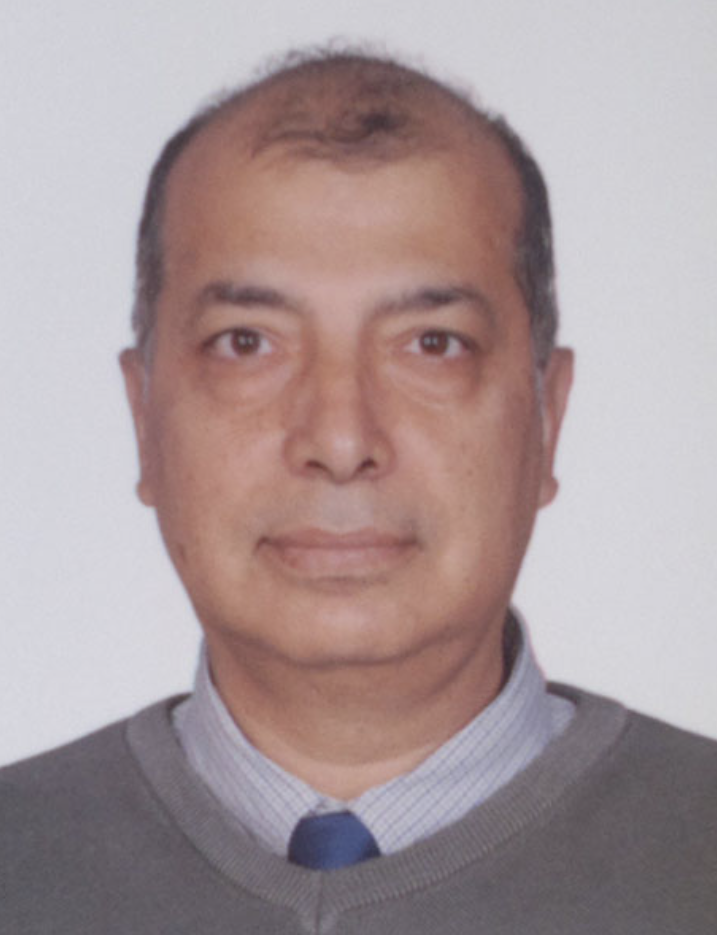}}]{Can Ozturan}
He received his Ph.D. degree in computer science from Rensselaer Polytechnic Institute, Troy, NY, USA, in 1995. After working as a Postdoctoral Staff Scientist at the Institute for Computer Applications in Science, NASA Langley Research Center, he joined the Department of Computer Engineering, Bogazici University in Istanbul, Turkey, as a faculty member in 1996. His research interests are blockchain technologies, parallel processing, scientific computing, resource management, graph algorithms and grid/cloud computing. He participated in SEEGRID2, SEEGRID-SCI, and PRACE 1IP, 2IP, and 3IP European FP7 infrastructure projects and is currently participating in Infinitech Horizon 2020 project.
\end{IEEEbiography}


\begin{thebibliography}{1}

\bibitem{nakamoto2008}  S. Nakamoto, ”Bitcoin: A peer-to-peer electronic cash system”, Decentralized Business Review, 21260, 2008.

\bibitem{schar2021} F.  Schär, "Decentralized finance: On blockchain-and smart contract-based financial markets", FRB of St. Louis Review, 2021.

\bibitem{vara2018}  R. Casado-Vara, J. Prieto, F. De la Prieta, J. M. Corchado, "How blockchain improves the supply chain: Case study alimentary supply chain", Procedia computer science, vol. 134, no. 393-398, 2018.

\bibitem{ali2020} F. S. Ali, O. Bouachir, Ö. Özkasap, M. Aloqaily, "SynergyChain: Blockchain-assisted adaptive cyber-physical P2P energy trading", IEEE Transactions on Industrial Informatics, vol. 17, no. 8, pp. 5769-5778, 2020.

\bibitem{yavuz2018} E. Yavuz, A. K. Koc, U. C. Cabuk, G. Dalkılıc, ”Towards secure e- voting using ethereum blockchain”, 6th International Symposium on Digital Forensic and Security (ISDFS), pp. 1-7, 2018.

\bibitem{ozturan2020}  C. Ozturan, ”Barter machine: an autonomous, distributed barter ex- change on the ethereum blockchain”, Ledger, vol. 5, 2020.

\bibitem{goldwasser1989}  S. Goldwasser, S. Micali, C. Rackoff, ”The knowledge complexity of interactive proof systems”, SIAM Journal on computing, vol. 18, no. 1, pp. 186-208, 1989.

\bibitem{cramer1994} R. Cramer, I. Damgard, B. Schoenmakers, ”Proofs of partial knowledge and simplified design of witness hiding protocols”, Annual International Cryptology Conference, pp. 174-187, 1994.

\bibitem{feige1988} U. Feige, A. Fiat, A. Shamir, ”Zero-knowledge proofs of identity”, Jour- nal of cryptology, vol. 1, no. 2, pp. 77-94, 1988.

\bibitem{guillou1988}  L. C. Guillou, J. J. Quisquater, ”A paradoxical indentity-based signature scheme resulting from zero-knowledge”, Conference on the Theory and Application of Cryptography, pp. 216-231, 1988.

\bibitem{fiat1986} A. Fiat, A. Shamir, "How to prove yourself: Practical solutions to identification and signature problems", In Conference on the theory and application of cryptographic techniques, pp. 186-194, 1986.

\bibitem{bellare1990} M. Bellare, S. Micali, R. Ostrovsky, "Perfect zero-knowledge in constant rounds", In Proceedings of the twenty-second annual ACM symposium on Theory of Computing, pp. 482-493, 1990.

\bibitem{goldreich1991} O. Goldreich, S. Micali, A. Wigderson, ”Proofs that yield nothing but their validity or all languages in NP have zero-knowledge proof systems”, Journal of the ACM (JACM), vol. 38, no. 3, pp. 690-728, 1991.

\bibitem{mohr2007}  A. Mohr, "A survey of zero-knowledge proofs with applications to cryptography", Southern Illinois University, Carbondale, pp. 1-12, 2007.

\bibitem{sasson2014}  E. Ben-Sasson, A. Chiesa, E. Tromer, M. Virza, ”Succinct Non-Interactive Zero Knowledge for a von Neumann Architecture”, 23rd USENIX Security Symposium (USENIX Security 14, pp. 781-796, 2014.

\bibitem{sasson2018} E. Ben-Sasson, I. Bentov, Y. Horesh, M. Riabzev, ”Scalable, transparent, and post-quantum secure computational integrity” Cryptology ePrint Archive, 2018.

\bibitem{bunz2019}  B. Bünz, J. Bootle, D. Boneh, A. Poelstra, P. Wuille, G. Maxwell, ”Bulletproofs: Short proofs for confidential transactions and more”, IEEE symposium on security and privacy (SP), pp. 315-334, 2018.

\bibitem{gabizon2019}  A. Gabizon, Z. J. Williamson, O. Ciobotaru, ”Plonk: Permutations over lagrange-bases for oecumenical noninteractive arguments of knowledge” Cryptology ePrint Archive, 2019.

\bibitem{gabay2020} D. Gabay, K. Akkaya, M. Cebe, "Privacy-preserving authentication scheme for connected electric vehicles using blockchain and zero knowledge proofs", IEEE Transactions on Vehicular Technology, vol., 69, no. 6, pp. 5760-5772, 2020.

\bibitem{li2020}  W. Li, H. Guo, M. Nejad, C. C. Shen, ”Privacy-preserving traffic man- agement: A blockchain and zero-knowledge proof inspired approach”, IEEE access, vol. 8, pp. 181733-181743, 2020.

\bibitem{jeong2021} S. Jeong, B. Ahn, "Implementation of real estate contract system using zero knowledge proof algorithm based blockchain", The Journal of Supercomputing, vol. 77, no. 10, pp. 11881-11893, 2021.

\bibitem{li2021} H. Li, W. Xue, ”A Blockchain-Based Sealed-Bid e-Auction Scheme with Smart Contract and Zero-Knowledge Proof”, Security and Communication Networks, 2021.

\bibitem{sha256} FIPS PUB 180-2, “SHA-256 Standard”, National Institute of Standards and Technology (NIST), 2002, (http://csrc.nist.gov/publications/fips/fips180-2/fips180-2withchangenotice.pdf).

\bibitem{eberhardt2018} J. Eberhardt, S. Tai, ”Zokrates-scalable privacy-preserving off-chain computations”, IEEE International Conference on Internet of Things (iThings) and IEEE Green Computing and Communications (Green- Com) and IEEE Cyber, Physical and Social Computing (CPSCom) and IEEE Smart Data (SmartData), pp. 1084-1091, 2018.

\bibitem{ecies} SECG SEC1, "Elliptic Curve Cryptography, Standards for Efficient Cryptography Group", ver. 2, 2009, (http://www.secg.org/download/aid-780/sec1-v2.pdf).

\bibitem{lindell2005} Y. Lindell, "Secure multiparty computation for privacy preserving data mining", In Encyclopedia of Data Warehousing and Mining, pp. 1005-1009). IGI global, 2005.

\bibitem{peng2019} F. Peng, S. Tang, B. Zhao, Y. Liu, "A privacy-preserving data aggregation of mobile crowdsensing based on local differential privacy", In Proceedings of the ACM Turing Celebration Conference, pp. 1-5, 2019.

\bibitem{singh2021} P. Singh, M. Masud, M. S. Hossain, A. Kaur, "Blockchain and homomorphic encryption-based privacy-preserving data aggregation model in smart grid", Computers \& Electrical Engineering, vol. 93, 107209, 2021.

\bibitem{almalki2021} F. A. Almalki, B. O. Soufiene, "EPPDA: an efficient and privacy-preserving data aggregation scheme with authentication and authorization for IoT-based healthcare applications", Wireless Communications and Mobile Computing, 2021.

\bibitem{fan2020} H. Fan, Y. Liu, Z. Zeng, "Decentralized privacy-preserving data aggregation scheme for smart grid based on blockchain" Sensors, vol. 20, no. 18, 5282, 2020.

\bibitem{mao2018}  T. Mao, C. Cao, X., Peng, W. Han, "A privacy preserving data aggregation scheme to investigate apps installment in massive mobile devices", Procedia Computer Science, vol. 129, pp. 331-340, 2018.

\bibitem{mols2020} J. Mols, E. Vasilomanolakis, "ethVote: Towards secure voting with distributed ledgers", In 2020 International Conference on Cyber Security and Protection of Digital Services (Cyber Security), pp. 1-8, 2020.

\bibitem{liu2023} Z. Liu, K. Zheng, L. Hou, H. Yang, K. Yang,  "A Novel Blockchain-Assisted Aggregation Scheme for Federated Learning in IoT Networks", IEEE Internet of Things Journal, 2023.

\bibitem{drogkaris2015} P. Drogkaris, A. Gritzalis "A privacy preserving framework for big data in e-government environments", In International Conference on Trust and Privacy in Digital Business, pp. 210-218, Springer, 2015.

\bibitem{damgard2012} I. Damgard, V. Pastro, N.Smart, S. Zakarias, "Multiparty computation from somewhat homomorphic encryption", In Annual Cryptology Conference, pp. 643-662, Springer, 2012.

\bibitem{damgard2013} I. Damgard, M. Keller, E. Larraia, V., Pastro, P., Scholl, N. P. Smart, "Practical covertly secure MPC for dishonest majority–or: breaking the SPDZ limits", In European Symposium on Research in Computer Security, pp. 1-18, Springer, 2013.

\bibitem{baldimtsi2015} F. Baldimtsi, O. Ohrimenko, "Sorting and searching behind the curtain", In International Conference on Financial Cryptography and Data Security, pp. 127-146, Springer, 2015.

\bibitem{bonawitz2017} K. Bonawitz, V. Ivanov, B. Kreuter, A. Marcedone, H. B. McMahan, S. Patel, et al., "Practical secure aggregation for privacy-preserving machine learning", In proceedings of the 2017 ACM SIGSAC Conference on Computer and Communications Security, pp. 1175-1191, 2017.

\bibitem{balle2020} B. Balle, J. Bell, A. Gascón, K. Nissim, "Private summation in the multi-message shuffle model", In Proceedings of the 2020 ACM SIGSAC Conference on Computer and Communications Security, pp. 657-676, 2020.

\bibitem{ranbaduge2020} T. Ranbaduge, D. Vatsalan, P. Christen, "Secure Multi-party Summation Protocols: Are They Secure Enough Under Collusion?", Trans. Data Priv., vol. 13, no. 1, pp. 25-60, 2020.

\bibitem{wang2021} N. Wang, S. C. K. Chau, Y. Zhou, "Privacy-preserving energy storage sharing with blockchain and secure multi-party computation", ACM SIGENERGY Energy Informatics Review, vol. 1, no. 1, pp. 32-50, 2021.

\bibitem{mouris2021} D. Mouris, N. G. Tsoutsos, "Masquerade: Verifiable Multi-Party Aggregation with Secure Multiplicative Commitments", Cryptology ePrint Archive, 2021.

\bibitem{menezes2018} A. J. Menezes, P. C. Van Oorschot, S. A. Vanstone, "Handbook of applied cryptography", CRC press, 2018.

\bibitem{blum1991} M. Blum, A. De Santis, S. Micali, G. Persiano, "Noninteractive zero-knowledge", \textit{SIAM Journal on Computing}, vol. 20, no. 6, pp. 1084-1118, 1991.

\bibitem{leighton2014}  F. T. Leighton,  "Introduction to parallel algorithms and architectures: Arrays, trees, hypercubes", Elsevier, 2014.

\bibitem{martinez2015} V. G. Martínez, E. H. Encinas, A. Q. Dios, "Security and practical considerations when implementing the elliptic curve integrated encryption scheme", Cryptologia, vol. 39, no. 3, pp. 244-269.

\bibitem{gilbert2003} H. Gilbert, H. Handschuh, "Security analysis of SHA-256 and sisters", In International workshop on selected areas in cryptography, pp. 175-193, Springer, 2003.

\bibitem{katseff1988} H. P. Katseff, "Incomplete hypercubes", IEEE transactions on computers, vol. 37, no. 5, pp. 604-608, 1988.

\bibitem{prabhala1992} V. K. Prabhala, "Algorithms for Incomplete Hypercubes", Master's Theses, Western Michigan University, 851, 1992.

\bibitem{solidity} Solidity Language, URL: https://docs.soliditylang.org/en/v0.8.15/, last accessed: 01 January 2023.

\bibitem{remix} Remix Compiler, URL: https://remix.ethereum.org, last accessed: 01 January 2023.

\bibitem{eccrpyto} Eccrypto Javascript Library, URL: https://www.npmjs.com/package/eccrypto, last accessed: 01 January 2023.

\bibitem{zokratesjs} Zokrates-js Javascript Library, URL: https://www.npmjs.com/package/zokrates-js, last accessed: 01 January 2023.

\bibitem{ethers} Ethers Javascript Library, URL: https://docs.ethers.io/v5/, last accessed: 01 January 2023.

\bibitem{metamask}  MetaMask, URL: https://metamask.io/, last accessed: 01 January 2023.

\bibitem{browserify} Browserify Bundler, URL: https://browserify.org, last accessed: 01 January 2023.

\bibitem{webpack} Webpack Bundler, URL: https://webpack.js.org, last accessed: 01 January 2023.

\bibitem{github} PVSS: Private Value Summation System, URL: https://github.com/GoshgarIsmayilov/PVSS, last accessed: 01 January 2023.

\bibitem{fuji} Avalanche Fuji Blockchain Network, URL: https://www.avax.network, last accessed: 01 January 2023.

\end{thebibliography}
\end{document}